\documentclass[11pt]{article}

\usepackage{calc}
\makeatletter
\@addtoreset{equation}{section}

\makeatother
\usepackage{feynmf}
\usepackage{graphicx}
\usepackage{amssymb,amsfonts,amsmath}

\def\beq{\begin{equation}}
\def\eeq{\end{equation}}
\def\bea{\begin{eqnarray}}
\def\eea{\end{eqnarray}}

\newcommand{\tup}{\theta^{(1)}_+}
\newcommand{\tdp}{\theta^{(2)}_+}
\newcommand{\tum}{\theta^{(1)}_-}
\newcommand{\tdm}{\theta^{(2)}_-}
\newcommand{\uphi}{\underline{\Phi}}
\newcommand{\ups}{\underline{\Psi_1}}
\newcommand{\upS}{\underline{\Psi_2}}

\begin{document}

\renewcommand{\thefootnote}{\fnsymbol{footnote}}
\newcommand{\inst}[1]{\mbox{$^{\text{\textnormal{#1}}}$}}
\begin{flushright}
DFTT 17/2011\\
EPHOU 11-004\\
November, 2011
\end{flushright}
\mbox{}\\\bigskip\bigskip
\begin{center}
{\LARGE Species Doublers as Super Multiplets
in Lattice Supersymmetry: \\
Chiral Conditions of Wess-Zumino Model for $D=N=2$ }\\[8ex]
%
{\large
Alessandro D'Adda\inst{a}\footnote{\texttt{dadda@to.infn.it}},
Issaku Kanamori\inst{b}
\footnote{\texttt{issaku.kanamori@physik.uni-regensburg.de}},
Noboru Kawamoto\inst{c}\footnote{\texttt{kawamoto@particle.sci.hokudai.ac.jp}}\\[1ex]
{\normalsize and}
Jun Saito\inst{c}\footnote{ \texttt{saito@particle.sci.hokudai.ac.jp}}. }\\[4ex]
%
{\large\itshape
\inst{a} INFN Sezione di Torino, and\\
Dipartimento di Fisica Teorica,
Universita di Torino\\
I-10125 Torino, Italy\\[3ex]
\inst{b}Institut f\"{u}r Theoretische Physik, Universit\"{a}t
Regensburg, \\
D-93040 Regensburg, Germany\\[3ex]
\inst{c} Department of Physics, Hokkaido University\\
Sapporo, 060-0810 Japan}
\end{center}
\bigskip\bigskip
\setcounter{footnote}{0}
\renewcommand{\thefootnote}{\arabic{footnote}}

\begin{abstract}
We propose an algebraic lattice supersymmetry formulation which
has an exact supersymmetry on the lattice.
We show how lattice version of chiral conditions can be
imposed to satisfy an exact lattice supersymmetry algebra.
The species doublers of chiral fermions and the corresponding
bosonic counterparts can be accommodated to fit into chiral
supermultiplets of lattice supersymmetry and thus lattice chiral
fermion problem does not appear. We explicitly show
how N=2 Wess-Zumino model in one and two dimensions can be
formulated to keep exact supersymmetry for all super charges on
the lattice. The momentum representation of $N=2$ lattice chiral
supersymmetry algebra has lattice periodicity and thus
momentum conservation should be modified to a lattice version of
sine momentum conservation, which generates nonlocal interactions
and leads to a loss of lattice translational invariance.
It is shown that the nonlocality is mild and the translational
invariance is recovered in the continuum limit.
In the coordinate representation a new type of product is defined
and the difference operator satisfies Leibnitz rule and an exact
lattice supersymmetry is realized on this product.
\end{abstract}




\section{Introduction}
\label{section1}

Although supersymmetry has not yet been discovered in nature, it plays a crucial role
 in various approaches to unified theories. It is thus very natural
to try to find a constructively well defined formulation of supersymmetric
theories working even in the nonperturbative regime. A lattice formulation
could be a good candidate. On the other hand it seems that the lattice
regularization and supersymmetry does not get along with each
other\cite{Montvay,Feo:2002yi,Kaplan:2003uh,Catterall:2004np,Giedt:1,CKU}.
Realization of exact lattice supersymmetry with a finite lattice spacing
has been a long standing problem since the first attempt of lattice
supersymmetry\cite{D-N}. It has never been realized in a satisfactory
way.

The lattice chiral fermion problem was also a longstanding problem but
it is considered that a satisfactory understanding was reached at
least for lattice QCD\cite{chiral-fermion}.
It is, however, not obvious that these two fundamental questions on the
lattice are not related. We consider that these two problems are related
in a fundamental way.

There are two major difficulties for constructing
a formulation of exact supersymmetry on the lattice.
Firstly the difference operator
which plays the role of differential operator on the lattice does
not satisfy the Leibnitz rule\cite{D-N,Fujikawa:Leibniz}.
A typical form of a lattice supersymmetry
algebra would have the form:  anti-commutator of two super charges is
equal to a difference operator.
However a supercharge operation on a product
of fields should satisfy  Leibnitz rule while the difference operator
does not satisfy the Leibnitz rule and thus there appears an
algebraic inconsistency already at the leading order of
lattice constant level.
Secondly there is the notorious chiral fermion problem. If we
naively put massless fermions on a lattice, species
doublers appear and an unbalance between the number of fermions
and bosons is generated. Thus supersymmetry breaking is
generated from different breaking source. This doubling of the chiral
fermion problem could, however, be avoided if we adopt other fermion
formulation with recently developed chiral fermion
approach\cite{chiral-fermion}.
There appear, however, other problems due to the non-equal footing
treatment of bosons and
fermions\cite{C-G,Giedt,Bergner:2009vg,KBUWW}.
In this paper we propose a formulation to solve the above
two difficulties.

As for the first difficulty it has been recognized in recent years
that the reason why we cannot
construct exact lattice supersymmetry models is not because it
is simply difficult but it is in fact impossible if we introduce
the difference operator in the lattice supersymmetry algebra
and keep strictly the locality and translational invariance on
the lattice\cite{Kato:2008sp,KatoSakamotoSo,Bergner:2009vg}.
If we give up some of the presumed characteristics such as locality,
translational invariance, commutativity,...
then we can find some supersymmetric lattice formulations which were
claimed to be exactly supersymmetric on the
lattice\cite{Nojiri1,BartelsKramer,KatoSakamotoSo,DKKN}.

We consider that there are three possibilities for exact lattice
supersymmetry formulations with a lattice derivative; \\
(1) Keeping a local lattice derivative like difference operator
but losing translational invariance on the lattice,  (2) keeping
finite translational invariance exact but using non-local lattice derivative
like SLAC derivative\cite{SLAC-der}, (3) keeping the difference operator as
lattice derivative operator but introducing non-commutativity between
fields, turning the  algebraic structure into a Hopf algebra.

The first possibility (1) was suggested by Dondi and Nicolai\cite{D-N} with
a lattice version of sine momentum conservation.
A possible realization of this idea was
later developed in the coordinate space\cite{Nojiri1}.
As a result it is equivalent to introduce a delta function of lattice
momentum, leading to nonlocal interactions\cite{D'Adda:2010pg}.
It has been recognized by many authors that the second possibility (2)
may be the only possibility if we strictly keep the lattice translational
invariance\cite{BartelsKramer,Kato:2008sp,KatoSakamotoSo,Bergner:2009vg}.
It is, however, well known that the SLAC type derivative
is highly nonlocal and may cause various problems.
In particular anomaly will not be generated properly when
gauge fields are turned on\cite{K-S2}, suggesting a fundamental connection
with the chiral fermion issue.
An alternative analysis by a consistency condition of block spin
transformation for supersymmetric lattice models led also to a statement
that the only consistent lattice derivative is the SLAC derivative\cite{BBP}.
It is important to realize that both  approaches (1) and (2) include
nonlocal interactions.
There was a proposal of exact supersymmetry without nonlocality but
with the introduction of infinite flavors\cite{KatoSakamotoSo}.

Our link approach formulation~\cite{DKKN} of lattice supersymmetry falls into
the case (3). According to the breakdown of the Leibnitz rule for
the difference operator we introduced similar breaking pattern of
Leibnitz rule for super charges and imposed algebraic consistency.
We then claimed that the exact lattice supersymmetry is
realized.
It was, however, pointed out that the formulation contains an
ordering ambiguity for the products of fields\cite{Dutch,B-C-K}.
It was later
recognized that the problems caused by the ordering ambiguity can
be solved by introducing a noncommutativity for fields carrying a
shift. Then it was claimed that the models of link approach are
exactly supersymmetric under a Hopf algebraic symmetry where the
difference operator can be treated as derivative operator
and the noncommutativity of fields
naturally fits within the scheme of a Hopf
algebra\cite{D'Adda:2009jb}.
In the case of gauge theory, the problem of gauge invariance was
questioned and remained to be understood.

We have so far considered the cases where the supersymmetry algebra
includes a lattice differential operator.
If a lattice differential operator is not included as part of the exact lattice
supersymmetry formulation, then the above arguments do not apply.
In fact exact lattice supersymmetry has been realized
in some special case of nilpotent super charge $Q^2=0$
which is a part  of an extended
supersymmetry algebra\cite{old-Dirac-Kahler,Kaplan:2003uh,Kaplan:2003,
Cohen:2003xe,catterall,sugino,DKKN,D-M,Takimi:2007nn,CKU}
in a twisted supersymmetry formulation\cite{KKU}.
The exactness of the nilpotent super charges of an extended
supersymmetry greatly helps the lattice models to converge
into supersymmetric continuum theories in lower
dimensions\cite{Kaplan:2003}, it is, however, expected that fine
tunings related
to other super charges may be difficult for higher dimensional
models with extended supersymmetry.

In our previous paper \cite{D'Adda:2010pg}, denoted in the following as
paper I, we proposed a new formulation
which has $N=2$ exact lattice supersymmetry in one dimension.
Our formulation, however, falls into the case (1) for the
first difficulty in the above classification. We introduced a
lattice version of sine
momentum conservation which led to nonlocal interactions
and a mild loss of translational invariance.
In  paper I we solved the second difficulty of species doubler
problem of chiral
fermions by identifying
 the species doublers of a chiral fermion  with members of its same
  supermultiplet of the $N=2$ supersymmetry.
The species doublers of chiral fermions are physical particles
and thus chiral fermion problem does not exist in this
formulation. We conjectured a possibility that the approach
(1) of  paper I and the link approach (3) are essentially equivalent.

In this paper we extend the previously developed formulation
of paper I to  the $N=2$ Wess-Zumino model in two dimensions. In particular
we develop a systematic treatment of chiral super fields
algebraically and find chiral conditions on the lattice.
Previously in paper I we found the supersymmetric action heuristically,
while here we provide a formulation for deriving the action as a
$Q$-exact quantity within the lattice version of chiral superfield formalism.

This paper is organized as follows:
We first explain the basic concept of lattice supersymmetry in
our formulation for the simplest model in section 2.
In section 3 we re-derive the action of one dimensional
$N=2$ supersymmetric model which was heuristically derived in paper I.
Superderivatives are consistently introduced on  the lattice and
the basic algebraic structure of lattice chiral superfield formulation are
presented for the one dimensional model.
In section 4 two dimensional full treatment of the $N=2$
lattice supersymmetry algebra is given. The chiral
conditions of the lattice supersymmetry algebra are derived.
In section 5 the Wess-Zumino action for $D=N=2$ is derived
in terms of lattice fields belonging to irreducible chiral
and antichiral representations.
In section 6 we introduce a new type of star product in
the coordinate representation on which
the difference operator satisfy the Leibnitz rule. We then
express the non local action in the coordinate space in terms of the star
product.
The section 7 includes conclusions and discussions.
In the appendix we provide an analysis of the recovery of
translational invariance in the continuum limit.

\section{The origin of lattice supersymmetry transformation}
\label{section2}

The simplest supersymmetry algebra in one dimension is given it terms of a single supercharge
$Q$ satisfying the following relation:
\beq
Q^2 = ~i \frac{\partial}{\partial x} \label{susyalg}  ~~,
\eeq
where  $i \frac{\partial}{\partial x}$
can be considered as the momentum generator. The lattice counterpart
of this algebra is obtained by replacing the differential
operator with  a difference operator
\beq
Q^2 = ~i \partial^{(s)} \label{latsusyalg}  ~~,
\eeq
where we introduce the symmetric difference operator\footnote{A symmetric
 operator is required to preserve hermiticity.}:
$\partial^{(s)}$:
\beq
\partial^{(s)}\Phi(x) = ~\frac{1}{a}\left\{\Phi\left(x+\frac{a}{2}\right)-
\Phi\left(x-\frac{a}{2}\right)\right\}  \label{symdiffop}  ~~,
\eeq
where $a$ is the fundamental lattice constant.

Since $i \partial^{(s)}$ can be considered as the generator
of a translation of the  lattice distance $a$, it is natural to expect that the super charge
$Q$ may be identified as a half lattice translation generator.
It is thus very natural to introduce half lattice sites to accommodate
the half lattice translation.
In our previous papers we stressed that the importance of the half
lattice structure in the lattice supersymmetry transformation can be
well understood by the matrix formulation of the super coordinates
and charges~\cite{Arianos:2008ai,D'Adda:2009es}.
Furthermore it was recognized that in the $D=N=1$ supersymmetry
 an alternating sign structure $(-1)^{\frac{2x}{a}}, ~(x=n\frac{a}{2},~ n \in Z)$ on the
half lattice is associated to the fermionic states.
One can then write a lattice ``superfield'' symbolically as
\beq
\Phi(x) = \varphi(x) + \frac{1}{2} \sqrt{a}(-1)^\frac{2x}{a} \psi(x)
\label{lattsup},
\eeq
where we have introduced a factor $\frac {1}{2}$ and $\sqrt{a}$ for
convenience. $\varphi$ and $\psi$ are, respectively, Grassman even and
odd fields.

Since we identify the super charge operation as a half lattice symmetric difference operation,
 one is led to introduce
lattice sites at integer multiples of $\frac{a}{4}$.
Then the expression (\ref{lattsup}) actually means:
\beq
\Phi(x) = \left\{ \begin{array}{lc} & \varphi(x)~~~~~~~~\textrm{for}~~~
 x=n a/2 ,\\&\frac{1}{2} a^{1/2} e^{\frac{2i \pi x}{a}}
 \psi(x) ~~~\textrm{for}~~~x=(2n+1)a/4 .
 \end{array} \right. \label{spf}
\eeq
The supersymmetry transformations can be identified as a half lattice
$\frac{a}{2}$ shift transformation of the superfield $\Phi(x)$:
\beq
 \delta_\alpha\Phi(x) = \alpha a^{-1/2} e^{\frac{2 i \pi x}{a}}
\left[ \Phi\left(x+\frac{a}{4}\right) -
\Phi\left(x-\frac{a}{4}\right) \right], \label{st}
\eeq
where $\alpha$ is a Grassman odd super parameter accompanying an alternating
sign. By separating $\Phi(x)$ into its component fields according to
 (\ref{spf}) we find:
\bea
&&\delta_\alpha \varphi(x) = \frac{i \alpha}{2} \bigg[
\psi\left(x+\frac{a}{4}\right) +
\psi\left(x-\frac{a}{4}\right) \bigg] \xrightarrow[a\to 0]{}
i \alpha \psi(x) \, ,\label{suslattf1} \\
&& \delta_\alpha \psi(x) =
2 a^{-1} \alpha \bigg[ \varphi\left(x+\frac{a}{4}\right)
- \varphi\left(x-\frac{a}{4}\right)  \bigg]
\xrightarrow[a\to 0]{}
\alpha
\frac{\partial \varphi(x)}{\partial x} \, , \label{suslattf2}
\eea
where $x$ is an even multiple of $a/4$ in (\ref{suslattf1}) and an
odd one in (\ref{suslattf2}).
It is rather surprising that the half lattice translation together with
alternating sign structure for the lattice superfields generates
a correct lattice supersymmetry transformation in the coordinate space.

The commutator of two  supersymmetry transformations (\ref{st})
leads  to a translation of the lattice constant $a$ generated
by the symmetric difference operator
$\partial^{(s)}$ given in eq.(\ref{symdiffop}), thus realizing the lattice
supersymmetry algebra (\ref{latsusyalg}):
\beq
 \delta_{\beta}\delta_{\alpha}\Phi(x) -
\delta_{\alpha}\delta_{\beta}\Phi(x) = 2\alpha\beta Q^2 \Phi(x) =
\frac{2 i \alpha \beta}{a}e^{\frac{4i\pi x}{a}}
\left[ \Phi\left(x+\frac{a}{2}\right)-
\Phi\left(x-\frac{a}{2}\right) \right]. \label{trs1}
\eeq
The exponential phase factor $e^{\frac{4i\pi x}{a}}$ is $+1$ for $x=\frac{na}{2}$ and $-1$ for
$x=\frac{na}{2}+\frac{a}{4}$.
The same algebraic relation (\ref{trs1}), but without the phase factor $e^{\frac{4i\pi x}{a}}$,
works separately also for the component fields $\varphi$ and $\psi$.

In  the coordinate space representation
the alternating sign structure plays a crucial role in reproducing
the correct lattice supersymmetry transformation and in interchanging the role
of the fermion and of the boson.
Let us consider the meaning of the alternating sign structure in the
momentum space representation. We define the Fourier transform of the fermionic
component field as
\beq
\psi(p) = \frac{1}{2}\sum_{x=\frac{n a}{2} + \frac{a}{4}} e^{ipx} \psi(x)
=- \psi\left(p\pm\frac{4\pi}{a}\right),
\label{fourier-psi}
\eeq
which has anti-periodicity with respect to a momentum shift of
$\frac{4\pi}{a}$,  since fermions are located on the quarter lattice sites.
Let us consider now  the fermion field with an accompanying alternating sign. In the
momentum space representation it becomes:
\beq
\frac{1}{2}\sum_{x=\frac{n a}{2} + \frac{a}{4}}
e^{ipx}(-1)^{\frac{2x}{a}}\psi(x)
= \psi\left(p+\frac{2\pi}{a}\right) = - \psi\left(p-\frac{2\pi}{a}\right).
\eeq
The alternating sign of the half lattice structure corresponds in  momentum space
 to a shift of $\pm\frac{2\pi}{a}$ in the momentum.
So the  fermion field with an accompanying  alternating sign structure in the
coordinate space may be associated to a species doubler field.

In a naive lattice formulation of chiral fermions it is known that the
species doublers appear as independent fields so that the number of degrees of freedom
of the fermions is increased and the matching of the degrees of freedom
between bosons and fermions may be lost.
However in our lattice formulation it is natural to introduce bosonic counter parts of species doublers
as well, thus restoring the balance between bosons and fermions.

In fact, since the lattice spacing for each component field is $\frac{a}{2}$ ( with a relative shift
of $\frac{a}{4}$ between the boson and the fermion field), but translational invariance is associated to shift
of the lattice spacing $a$, for each field on the lattice there are two translational invariant
configurations, namely:
\beq
\Phi_1(x) = c_1, ~~~~~~\Phi_2(x) = (-1)^n c_2,
\label{const-states}
\eeq
where $\Phi \equiv \{\varphi, \psi \}$,  $x=\frac{na}{2}$ for $\Phi\equiv\varphi$ and
$x=\frac{na}{2}+\frac{a}{4}$ for $\Phi\equiv\psi$.

These two configurations, being translational invariant, should have zero momentum
in the continuum limit but on the lattice they have respectively momentum $0$ and $\frac{2\pi}{a}$.
Fluctuations around $\Phi_1$ and $\Phi_2$ describe distinct degrees of freedom, which in the case of fermions can be described as the field and its doubler.

So, although we started from an $N=D=1$ supersymmetry we end up with enough degrees of freedom
(including the doublers in the count) to accommodate an $N=2$ supersymmetry. This was accomplished in ref~\cite{D'Adda:2010pg}, and will be reviewed in the next section.

To summarize, the basic lattice supersymmetry structure as it emerges from the present discussion is:\\
{\it The lattice supersymmetry transformation is a half lattice
translation which mixes in general fields and doubler-fields, namely smooth fields on the
 $\frac{a}{2}$ lattice and fields with an alternating sign structure. The fermion doubling problem
 is solved as  doublers and original fields alike  are members of the same extended supersymmetry multiplet.} \\

\section{Revisit to one-dimensional $N=2$ model with lattice chiral conditions }
\label{section3}
\setcounter{equation}{0}

In this section we shall give a derivation of the lattice action invariant under
the $N=2$ supersymmetry in $1$ dimension.
This action was found in our  previous
paper \cite{D'Adda:2010pg} in a more heuristic way. We shall also derive the lattice version of chiral
conditions which play an important part in the derivation of the action.
This formulation will be extended to the
 two dimensional $N=2$ Wess-Zumino
model in the following  sections.

There are two supercharges in the $N=2$ model, whose algebra is given
as follows:
\beq
Q_1^2 = Q_2^2 = i\partial, ~~~~~~~~\{Q_1,Q_2\} =0,
\label{1d-N2-alg}
\eeq
which shows that there are two uncorrelated $N=1$ algebras of the type (\ref{susyalg}).
One of the  supersymmetry transformations, which we denote by
$\delta_1=\alpha Q_1$, can be identified with (\ref{suslattf1}) and
(\ref{suslattf2}) of the  $N=1$ model:
\bea
&&Q_1 \Phi(x) = \frac{i }{2} \bigg[
\Psi\left(x+\frac{a}{4}\right) + \Psi\left(x-\frac{a}{4}\right)
\bigg]~~~~~~~~x=\frac{n a}{2} \, ,
 \nonumber \\
&& Q_1 \Psi(x) = 2  \bigg[ \Phi\left(x+\frac{a}{4}\right)
- \Phi\left(x-\frac{a}{4}\right)  \bigg] ~~~~~~~~~x=\frac{n a}{2} +
\frac{a}{4} \, .
\label{Q1b-f} \eea
We assume here that $\Phi(x)$ and $\Psi(x)$ are dimensionless, so that no dependence on the lattice spacing $a$ appears at the r.h.s. of (\ref{Q1b-f}).
Of course a rescaling of the fields with powers of $a$ will be needed to make
contact with the fields of the continuum theory~\cite{D'Adda:2010pg}.
Let us introduce now the fields in  momentum space defined as the
Fourier transform of $\Phi(x)$ and $\Psi(x)$:
\beq
 \Phi(p) = \frac{1}{2} \sum_{x=\frac{n a}{2}} e^{ipx} \Phi(x), \, ~~~~~~~~~~~~~~~~~~\,
 \Psi(p) = \frac{1}{2} \sum_{x=\frac{n a}{2} + \frac{a}{4}} e^{ipx} \Psi(x). \, \label{ftrans}
\eeq

The corresponding inverse transformations are:
 \beq
 \Phi(x) = a \int_0^{\frac{4\pi}{a}} \frac{d p}{2\pi} \Phi(p)e^{-ipx},\, ~~~~~~~~~~~~~~~~~~\,
\Psi(x) = a \int_0^{\frac{4\pi}{a}} \frac{d p}{2\pi}
\Psi(p)e^{-ipx}. \, \label{invtrans} \eeq From(\ref{ftrans}) it is
clear that $\Phi(p)$ and $\Psi(p)$ satisfy the following
periodicity conditions: \beq \Phi\left(p+\frac{4 \pi}{a} \right) = \Phi(p),
\,~~~~~~~~~~~~~~~~~~\Psi\left(p+\frac{4 \pi}{a} \right) = - \Psi(p).
\label{period} \eeq In momentum representation the supersymmetry
transformations (\ref{Q1b-f}) read: \bea
 &&Q_1 \Phi(p) =  i \cos \frac{a p}{4}  \Psi(p) \, ,
\nonumber \\
&& Q_1 \Psi(p) = -4 i \sin \frac{a p}{4}   \Phi(p) \, .
\label{Q1pb-f}
\eea

We have now to identify the second supersymmetry transformation $Q_2$.
As discussed in~\cite{D'Adda:2010pg}, by comparing with the momentum representation of the continuum formulation
we find that the second lattice supersymmetry transformation can be obtained by doing in (\ref{Q1pb-f}) the
following replacement  for the fermion field:
  \beq
  \Psi(p) \longrightarrow -i \Psi\left(\frac{2\pi}{a}-p\right),  \label{replacement}
  \eeq
  obtaining for  $Q_2$ the following expression:
  \bea
&&Q_2 \Phi(p) =   \cos \frac{a p}{4}  \Psi\left(\frac{2\pi}{a}-p\right) \, ,
 \nonumber \\
&& Q_2 \Psi\left(\frac{2\pi}{a}-p\right) = 4  \sin \frac{a p}{4}   \Phi(p) \, .
\label{Q2pb-f} \eea

From (\ref{Q1pb-f}) and (\ref{Q2pb-f}) one can check that the  super charges
$Q_1$ and $Q_2$ satisfy on the lattice  in
momentum representation  the following algebra:
\beq
Q_1^2=Q_2^2=2\sin\frac{ap}{2}, ~~~~~~~~~~\{Q_1,Q_2\}=0.
\label{scharge-alg}
\eeq
This is the lattice momentum version of the algebra (\ref{1d-N2-alg}).

We introduce now the chiral  super charges
\beq
Q_{\pm}=\frac{1}{2}(Q_1\pm i Q_2),
\label{chiral-scharge}
\eeq
which satisfy the following chiral algebra:
\beq
\{Q_+,Q_-\}=2\sin\frac{ap}{2}, ~~~~~~~~~~Q_+^2=Q_-^2=0.
\label{chiral-salgebra-Q}
\eeq

We want to introduce the  lattice version of the superderivatives.
In the superspace formalism of the continuum theory the superderivative differential
 operators are obtained from  the corresponding supercharge operators by changing the
 sign of $\frac{\partial}{\partial x^\mu}$, namely  of the momentum.
In our lattice formulation where we have species doublers in the
momentum representation of the half lattice structure, the correspondence
between the lattice momentum and continuum momentum need to be carefully
defined. We make a correspondence with the lattice momentum $p$ and
continuum momentum $p_c$ in the following way:
\bea
&&(1)~ \hbox {Field fluctuations around }~ p = 0;~~~ p = p_c,
\nonumber \\
&&(2)~ \hbox {Species doublers fluctuations around}~ p = \frac{2\pi}{a};
~~~p=\frac{2\pi}{a} - p_c.
\label{cont-mom}
\eea

In the lattice momentum region (1), the sign change of momentum
$p_c\rightarrow -p_c$ corresponds to the following change of the coefficients
in the supersymmetry transformation:
\beq
\cos\frac{ap}{4}~\rightarrow ~\cos\frac{ap}{4},~~~
\sin\frac{ap}{4}~\rightarrow ~-\sin\frac{ap}{4},
\label{0-region}
\eeq
while in the region (2) the sign change of momentum
$p_c\rightarrow -p_c$ corresponds to
\bea
\cos\frac{ap}{4}~=~
\cos\frac{a}{4}\left(\frac{2\pi}{a}-p_c \right)~&\rightarrow& ~
\cos\frac{a}{4}\left(\frac{2\pi}{a}+p_c \right)~=~
-\cos\frac{ap}{4}
\nonumber \\
\sin\frac{ap}{4}~=~
\sin\frac{a}{4}\left(\frac{2\pi}{a}-p_c \right)~&\rightarrow& ~
\sin\frac{a}{4}\left(\frac{2\pi}{a}+p_c \right)~=~
\sin\frac{ap}{4}.
\label{spec-region}
\eea
Thus the sign changes of $\cos\frac{ap}{4}$ and $\sin\frac{ap}{4}$
are opposite in the region (1) and (2).
To connect smoothly eq.~(\ref{0-region}) to eq.~(\ref{spec-region}) one can  multiply
the right hand sides by a factor $\cos\frac{ap}{2}$, which just provides a $-1$ factor
at $p=\frac{2\pi}{a}$.
Hence to obtain the lattice superderivative from the supersymmetry transformation we simply do the following replacements:
\beq
\cos\frac{ap}{4}~\rightarrow~\cos\frac{ap}{2}\cos\frac{ap}{4},~~~
\sin\frac{ap}{4}~\rightarrow~-\cos\frac{ap}{2}\sin\frac{ap}{4}.
\label{change-to-susyder}
\eeq
The explicit form of the $N=2$ superderivative  in one dimension acting on the component fields
is:
\bea
 &&D_1 \Phi(p) =  i \cos \frac{a p}{2} \cos \frac{a p}{4}  \Psi(p) \, ,~~~
D_2 \Phi(p) =  \cos \frac{a p}{2} \cos \frac{a p}{4}
\Psi\left(\frac{2\pi}{a}-p\right) \, ,
\nonumber \\
&& D_1 \Psi(p) = 4 i\cos \frac{a p}{2} \sin \frac{a p}{4}   \Phi(p) \, ,~~~
D_2 \Psi\left( p\right) = 4  \cos \frac{a p}{2} \cos \frac{a p}{4}
\Phi\left(\frac{2\pi}{a}-p\right) \,
\label{su-cov-der-1d}
\eea
which satisfy the following algebra:
\bea
D_1^2 ~ = ~ D_2^2 ~ &=& ~ -2\cos^2\frac{ap}{2} \sin\frac{ap}{2}, ~~~~~
\{D_1,D_2\}~ = ~ 0,
\label{su-cov-alg-1d} \\
\{Q_i,D_j\} ~&=&~ 0 ~~~(i,j = 1,2).
\label{anti-com-QD}
\eea

Notice that the factor $\cos\frac{ap}{2}$, while essential to establish the
anticommuting nature of $Q_i$ and $D_i$, does not change the algebraic structure
in the continuum limit.

In analogy to chiral supercharges chiral superderivatives are defined as:
\beq
D_\pm ~ = ~ \frac{1}{2}(D_1 ~ \pm ~ i D_2),
\label{chi-sucov-der}
\eeq
which satisfy the chiral algebra together with (\ref{chiral-salgebra-Q}),
\bea
D_+^2 ~ = ~ D_-^2  &=& 0, ~~~\{D_+,D_-\}~=~-2\cos^2\frac{ap}{2} \sin\frac{ap}{2}\nonumber \\
\{Q_\pm,D_\pm\}&=&\{Q_\pm,D_\mp\}~=~0.
\label{chi-sucov-alg}
\eea

 Chiral conditions on the lattice can now be imposed:
\bea
D_-\Phi(p)&=&i\cos \frac{a p}{2} \cos \frac{a p}{4} \frac{1}{2}
\left\{ \Psi(p) ~-~\Psi\left(\frac{2\pi}{a} - p \right)\right\}~=~0,
\label{chi-cond-1} \\
D_-\Psi(p)&=&  i\sin ap ~~\frac{1}{2}
\left\{\frac{\Phi(p)}{\cos \frac{ap}{4}} -
\frac{\Phi\left(\frac{2\pi}{a} - p\right)}{\sin \frac{ap}{4}}
\right\} ~=~0.
\label{chi-cond-2}
\eea
The first chiral condition (\ref{chi-cond-1}) simply  states that
\beq
\Psi(p) = \Psi(\frac{2\pi}{a} - p )
\label{ch1}
\eeq
namely that the doublers degree of freedom is the same as the original fermion field.
The second condition (\ref{chi-cond-2}) has a similar interpretation in terms of the rescaled field
\beq
\uphi(p) \equiv \frac{\Phi(p)}{\cos \frac{ap}{4}},
\label{rescaled-phi}
\eeq
namely:
\beq
\uphi(p)~ = ~ \uphi\left(\frac{2\pi}{a} - p\right) .
\label{chi-cond-res}
\eeq
The antichiral conditions are obtained by imposing $D_+\Phi(p)=D_+\Psi(p)=0$ and
lead to similar conditions, but with a minus sign at the r.h.s. of (\ref{ch1}) and of
(\ref{chi-cond-res}). So if we decompose $\uphi$ and $\Psi$ as:
\beq
\uphi(p) = \uphi^{(c)}(p) +  \uphi^{(a)}(p),~~~~~~\Psi(p)= \Psi^{(c)}(p) + \Psi^{(a)}(p)
\label{dcps}
\eeq
with
\begin{eqnarray}
2\uphi^{(c)}(p) &=& \uphi(p) + \uphi\left(\frac{2\pi}{a} - p\right), ~~~~~
2\uphi^{(a)}(p) = \uphi(p) - \uphi\left(\frac{2\pi}{a} - p\right),
\nonumber \\
2\Psi^{(c)}(p) &=& \Psi(p) + \Psi\left(\frac{2\pi}{a} - p\right), ~~~~~
2\Psi^{(a)}(p) = \Psi(p) - \Psi\left(\frac{2\pi}{a} - p\right).
\label{sym-asym-fields}
\end{eqnarray}
$\uphi^{(c)}$ and $\Psi^{(c)}$ are the components of a chiral superfield on the
lattice, whereas $\uphi^{(a)}$ and $\Psi^{(a)}$ are the components of an antichiral superfield.
So eq. (\ref{dcps}) gives the decomposition of $\uphi$ and $\Psi$ into its irreducible chiral
and antichiral representations of the supersymmetry algebra.

The supersymmetry transformations of the chiral and anti-chiral fields
are given in Table 1, which also shows clearly that each of them form an irreducible
representation of the algebra.
Notice that due to the definition (\ref{rescaled-phi}) both $\uphi(p)$ and $\Psi(p)$ change sign
when $p$ is shifted of $\frac{4\pi}{a}$. In coordinate space this means that they are both located on odd integer multiples of $\frac{a}{4}$.
\begin{table}
\hfil
$
\begin{array}{|l|c|c|} \hline
   &\displaystyle  Q_+ & \displaystyle Q_-  \\ \hline
\hline
\uphi^{(c)}(p) & i\Psi^{(c)}(p) & 0 \\
\Psi^{(c)}(p) & 0 & -2i\sin \frac{ap}{2} \uphi^{(c)}(p) \\
\uphi^{(a)}(p) & 0 & i \Psi^{(a)}(p) \\
\Psi^{(a)}(p) & -2i\sin \frac{ap}{2} \uphi^{(a)}(p) & 0 \\ \hline
\end{array}
$
\caption{Chiral and anti-chiral $N=2$ supersymmetry transformation }
\end{table}
However one can see from Table 1 that the supersymmetry transformations are also consistent with
$\uphi^{(c)}(p)$ and $\Psi^{(c)}(p)$ being both periodic in $p$ ( same for
 $\uphi^{(a)}(p)$ and $\Psi^{(a)}(p)$). This can be achieved by simply shifting the origin of the
 coordinate space by $\frac{a}{4}$, in particular we could have defined from the very beginning
 the fermionic field $\Psi(x)$ on integer multiples of $\frac{a}{2}$ and the bosonic field
 $\Phi(x)$ on the half integer ones.

 However a shift of the origin is not without consequences with respect to the symmetry under
 $p \to \frac{2\pi}{a} -p$. In fact a shift of $\frac{a}{4}$ in $x$ amounts to a multiplication by
  $e^{i\frac{ap}{4}}$ in momentum space, and this is not invariant under $p \to \frac{2\pi}{a} -p$.

  If we assume, consistently with the definition of chiral and antichiral superderivatives, that chiral and antichiral superfields are hermitian conjugate, then one can prove that antiperiodicity in the momentum implies that chiral and antichiral fields have opposite symmetry with respect to $p \to \frac{2\pi}{a} -p$
  (as in the example above), whereas periodicity in the momentum requires that chiral and antichiral fields have the same symmetry with respect to $p \to \frac{2\pi}{a} -p$ (both symmetric or both antisymmetric).

  So both choices are ultimately allowed and this arbitrariness will eventually come useful in building the action, particularly in the more complex two dimensional case.
We want now to re-derive the action for an $n$-point interaction, action that in ref.~\cite{D'Adda:2010pg} was essentially postulated.

  In momentum representation the $n$-point interaction term will be expressed as an integral over the $n$ momenta of the fields.
  Each integration will be over a period of $\frac{4\pi}{a}$, so a condition to be required for the action density is to
  be periodic of period $\frac{4\pi}{a}$ in each momentum. Moreover consider, for each momentum $p_i$, the reflection  symmetry with respect to the point  $\frac{2\pi}{a}$, namely the symmetry  $p_i \to \frac{2\pi}{a} - p_i$, is required. It is easy to see that, being the integration interval $(-\frac{\pi}{a}, \frac{3\pi}{a})$ symmetric with respect to the point  $\frac{2\pi}{a}$,  the odd (antisymmetric) part of the action density vanishes identically.

  A supersymmetric invariant action can be written using the $Q$-exact expression
  \beq
   \mathcal{L}(p_1,p_2,\cdots,p_n) = -\frac{1}{n} \delta\left(\sin\frac{ap_1}{2}+\cdots+\sin\frac{ap_n}{2}\right) Q_+Q_- \left[ \Phi(p_1)\Phi(p_2)\cdots\Phi(p_n)\right] \label{invden}
  \eeq
  The right hand side in (\ref{invden}) can be calculated using the explicit
form of supersymmetry transformations, and gives:
  \bea
  &&\mathcal{L}(p_1,p_2,\cdots,p_n) = \delta\left(\sin\frac{ap_1}{2}+\cdots+\sin\frac{ap_n}{2}\right) \label{invden2} \\
  &&\{2 \sin^2\frac{a p_1}{4} ~\Phi(p_1)\Phi(p_2)\cdots\Phi(p_n) +
  \frac{n-1}{4}  \sin\frac{a(p_1-p_2)}{4} \Psi(p_1)\Psi(p_2)\Phi(p_3)
\cdots\Phi(p_n)\} \nonumber
\eea
which coincides with the expression given in ref.~\cite{D'Adda:2010pg} .
The invariance of $\mathcal{L}(p_1,p_2,\cdots,p_n)$ under supersymmetry transformation follows immediately from the supersymmetry algebra
(\ref{chiral-salgebra-Q}) and the sine momentum conservation. According to our original choice $\Phi(p)$ is periodic in $p$ with period
$\frac{4\pi}{a}$ and so this property is extended to  $\mathcal{L}(p_1,p_2,\cdots,p_n)$.
So the invariant action can be written as:
\beq
S_n = \int_{-\frac{\pi}{a}}^{\frac{3\pi}{a}} dp_1dp_2\cdots dp_n \left(\prod_{j=1}^n \cos\frac{ap_j}{2}\right) \mathcal{L}(p_1,p_2,\cdots,p_n)
\label{invaction}
\eeq
where the product of cosines has been introduced, as in reference~\cite{D'Adda:2010pg}, to compensate singularities of the integration volume at $p_i=\pm \frac{\pi}{a}$ arising from the $\delta$-function.
Notice that, according to what mentioned earlier about the symmetry under $p_i \to \frac{2\pi}{a} - p_i$, only the symmetric part of $\Phi(p_i)$, namely $\frac{1}{2}(\Phi(p_i) + \Phi(\frac{2\pi}{a}-p_i))$  contributes to the action. However $\Phi$ is not the normalized field $\uphi$, and its symmetric part is not a pure chiral field but contains a mixture of chiral and anti-chiral components.
This is different from what happens in the $N=D=2$ case, where the action will be constructed using chiral and anti-chiral fields.

\section{$N=2$ $D=2$ chiral superfields on the lattice}

In this section we shall generalize the previous approach to a two dimensional
$N=2$ supersymmetry algebra.
In extending the structure of the one dimensional lattice
supersymmetry transformation into two dimensions, it is important
to realize how one dimensional supersymmetry algebra is included
in the two dimensional algebra.
$N=2$ extended supersymmetry algebra in two dimensions is given by
\beq
\{Q_{\alpha i}, Q_{\beta j}\}~ = ~2\delta_{ij}
(\gamma^\mu)_{\alpha \beta} i\partial_\mu ,
\label{D=N=2-alg1}
\eeq
where we may use an explicit representation of Pauli matrices for
$\gamma^\mu=\{\sigma^3,\sigma^1\}$. By going to the light cone
directions this two dimensional $N=2$
algebra can be decomposed into the direct sum of two one dimensional $N=2$ algebra :
\beq
\{Q_\pm^{(i)},Q_\pm^{(j)}\} ~=~ 2\delta^{ij} i\partial_\pm, ~~~~~~
\{\hbox{others}\} =0,
\label{D=N=2-alg2}
\eeq
where
\beq
Q_\pm^{(j)}~=~\frac{Q_{1j}\pm i Q_{2j}}{\sqrt{2}}, ~~~~~~
\partial_\pm = \partial_1 \pm i\partial_2 ,
\label{def-scharge-N=D=2-1}
\eeq
where in terms of euclidean light cone coordinates
\beq
x_\pm=x_1\pm ix_2,~~~~~~  \partial_\pm=\frac{\partial}{\partial x_\pm}.
\label{euclid-light-cone}
\eeq
We can also equivalently express it in a chiral form:
\beq
\{Q_+^{(+)},Q_+^{(-)}\} ~=~i\partial_+, ~~~
\{Q_-^{(+)},Q_-^{(-)}\} ~=~i\partial_-, ~~~\{\hbox{others}\}=0,
\label{D=N=2-alg3}
\eeq
where
\beq
Q_\pm^{(+)} = \frac{Q_\pm^{(1)} + iQ_\pm^{(2)}}{2}, ~~~
Q_\pm^{(-)} = \frac{Q_\pm^{(1)} - iQ_\pm^{(2)}}{2}.
\label{def-scharge-N=D=2-2}
\eeq

Finally we can introduce the Dirac-K\"ahler twisting procedure by defining
twisted super charges\cite{KKU,DKKN}:
\beq
Q_{\alpha i} =
(Q{\bf 1} + Q_\mu \gamma^\mu +\gamma^2\gamma^1\tilde{Q})_{\alpha i},
\label{D=N=2-alg4}
\eeq
that  satisfy the following twisted supersymmetry algebra:
\beq
\{Q,Q_\mu\} = i\partial_\mu, ~~~~~~
\{\tilde{Q},Q_\mu \} = \epsilon_{\mu\nu} i\partial_\nu.
\eeq
 The twisted super charges are related to the chiral super charges by:
\bea
Q_+^{(+)} = \frac{Q_1 + iQ_2}{\sqrt{2}} ,&& ~~
Q_-^{(-)} = \frac{Q_1 - iQ_2}{\sqrt{2}},
\nonumber \\
Q_+^{(-)} = \frac{Q + i\tilde{Q}}{\sqrt{2}} ,&& ~~
Q_-^{(+)} = \frac{Q - i\tilde{Q}}{\sqrt{2}}.
\eea

As we can see from (\ref{D=N=2-alg2}), we
basically  introduce the one dimensional lattice
supersymmetry structure in each light cone direction to construct the
$D=N=2$ lattice supersymmetry formulation. We shall show  later that the chiral
version of algebra (\ref{D=N=2-alg3}) plays an important role in constructing chiral
and anti-chiral irreducible representations of the  algebra on the lattice.

In analogy with the one dimensional case we introduce
in each light cone direction  the one dimensional half lattice structure.
In order to introduce a lattice structure on the light cone coordinates we
need$ x_\pm$ to be real quantities. Therefore we have to go from Euclidean to
Minkowsky metric, namely, replace in (\ref{euclid-light-cone}) $ix_2$ with
$ x_2$. So in the following we shall refer to the light cone coordinates
$x_\pm$ as the ones in Minkowsky space, namely $x_\pm=x_1 \pm x_2$.

The half lattice sites on the light cone
coordinate can then be represented by
\beq
\vec{x}_\pm = (n\frac{a}{2},~ m\frac{a}{2}), ~~~~~~n,m\in {\bf Z},
\label{l-con-coor}
\eeq
where $a$ is the lattice spacing associated to an elementary translation. As in the one dimensional case
shifts of $\frac{a}{2}$ will be associated to supersymmetry transformations.
The lattice is illustrated in  Fig.~\ref{fig:lattice}. It is composed by the
union of four lattices each
  with spacing $a$ in the light cone directions. These lattices are represented in the figure by different types of dots.
The lattice with black dots on the sites represents points of the form $(x_+,x_-)=(n'a,m'a)$
with $n'=\frac{n}{2}$ and $m'=\frac{m}{2}$ both integers. In the other lattices the integers
$(n,m)$ are as follows: (odd,odd) on the sites with white dots, (odd, even) on the sites with crosses and (even,odd) on the sites with circular crosses.
  For instance the lattice with black dots on the sites represent points with $m$ and $n$ in (\ref{l-con-coor}) both even,
  namely points of the form $(x_1,x_2)=(n'a,m'a)$ with $n',m'\in {\bf Z}$. Similarly the lattice with white dots has the structure
  $(x_1,x_2)=(n'a+\frac{a}{2},m'a+\frac{a}{2})$, and so on.
  Points belonging to the same lattice can be reached from each other by shifts which are integer multiples of $a$ in each light cone direction,
  namely they can be reached by translations. Points belonging to different lattices, for instance a black dot site and a white dot site,
  cannot be reached from each other by translations as shifts which are odd multiples of $\frac{a}{2}$ are involved.

  As a result of the existence in our lattice of four distinct translational invariant sublattices, there are four distinct translational invariant field configurations, namely:

\begin{figure}
 \hfil\includegraphics[height=0.6\linewidth]{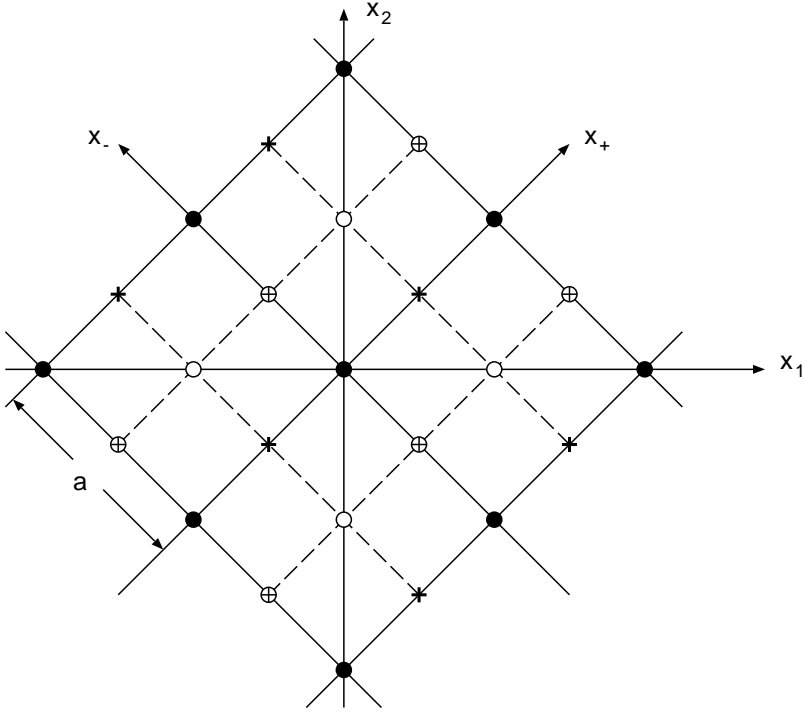}
 \caption{Light cone coordinate versus Cartesian coordinate }
 \label{fig:lattice}
\end{figure}

\beq
\Phi_1(\vec{x}) = c_1, ~~~~~~\Phi_2(\vec{x}) = (-1)^{n} c_2,~~~~~~\Phi_3(\vec{x}) = (-1)^{m} c_3,~~~~~~\Phi_4(\vec{x}) = (-1)^{n+m} c_4 ,
\label{const-states2}
\eeq
where the $c_i$ s  are constants and $n$ and $m$ refer to (\ref{l-con-coor}).
So for instance $\Phi_1(\vec{x})$ is $c_1$ on all sites, but $~\Phi_2(\vec{x})$ is $c_2$
on the sites with black dots and circular crosses and $-c_2$ on the others, and so on.

Considering  that physical fields are small fluctuations around a constant, i.e. translational invariant, configuration, it is clear from the previous discussion that a lattice field corresponds now to four physical degrees of freedom.  These can be  naturally identified with species doubler states, and this phase reminds us of the staggered phase of staggered fermion\cite{K-S}.

The algebra (\ref{D=N=2-alg2})  has $4$ supercharges, hence an $N=2$ superfield depends on $4$ odd Grassmann variables $\theta^{(i)}_\pm$ and its expansion contains $16$ component fields.
As in the one dimensional case we shall proceed by considering first the $N=1$ case, and by giving a realization of the $N=1$ superalgebra on a two dimensional lattice with spacing $\frac{a}{2}$. Since an $N=1$ superfield in $D=2$ contains $4$ component fields ($2$ bosonic and $2$ fermionic), the lattice realization of the $N=1$ superalgebra is done in terms of four fields on the lattice. However, because of the $\frac{a}{2}$ spacing, each field on the lattice comes together with three ``doublers'' and eventually corresponds to four degrees of freedom in the continuum for a total of $16$ degrees of freedom. We will then show that these $16$ degrees of freedom provide a representation on the lattice of the $N=2$ superalgebra, with the doublers of a field being different members of the same supermultiplet.

\subsection{$N=1$,$D=2$ superalgebra on the lattice}

Consider the $N=1$ subalgebra of (\ref{D=N=2-alg2}) obtained by setting, for instance, $i=j=1$. The superfield expansion in this case is given by:
\beq
\Phi(x_+,x_-,\theta_+^{(1)},\theta_-^{(1)}) = \varphi(x_+,x_-) + i \theta_+^{(1)} \psi_1(x_+,x_-) +i \theta_-^{(1)} \psi_2(x_+,x_-) + i \theta_-^{(1)} \theta_+^{(1)} F(x_+,x_-),
\label{sup-N1}
\eeq
 and the supersymmetry transformations are:
\begin{eqnarray}
\delta^{(1)}_+\varphi = i \alpha_+ \psi_1, ~~~~~
\delta^{(1)}_+\psi_1 = \alpha_+ \frac{\partial\varphi}{\partial x_+},
\nonumber \\
\delta^{(1)}_+\psi_2= \alpha_+ F, ~~~~~
\delta^{(1)}_+ F = i \alpha_+ \frac{\partial\psi_2}{\partial x_+}.
\label{delta+}
\end{eqnarray}
and
\begin{eqnarray}
\delta^{(1)}_-\varphi = i \alpha_- \psi_2, ~~~~~
\delta^{(1)}_-\psi_2 = \alpha_- \frac{\partial\varphi}{\partial x_-},
\nonumber \\
\delta^{(1)}_-\psi_1= -\alpha_- F, ~~~~~
\delta^{(1)}_- F = -i \alpha_- \frac{\partial\psi_1}{\partial x_-}.
\label{delta-}
\end{eqnarray}

\begin{figure}
 \hfil\includegraphics[height=0.6\linewidth]{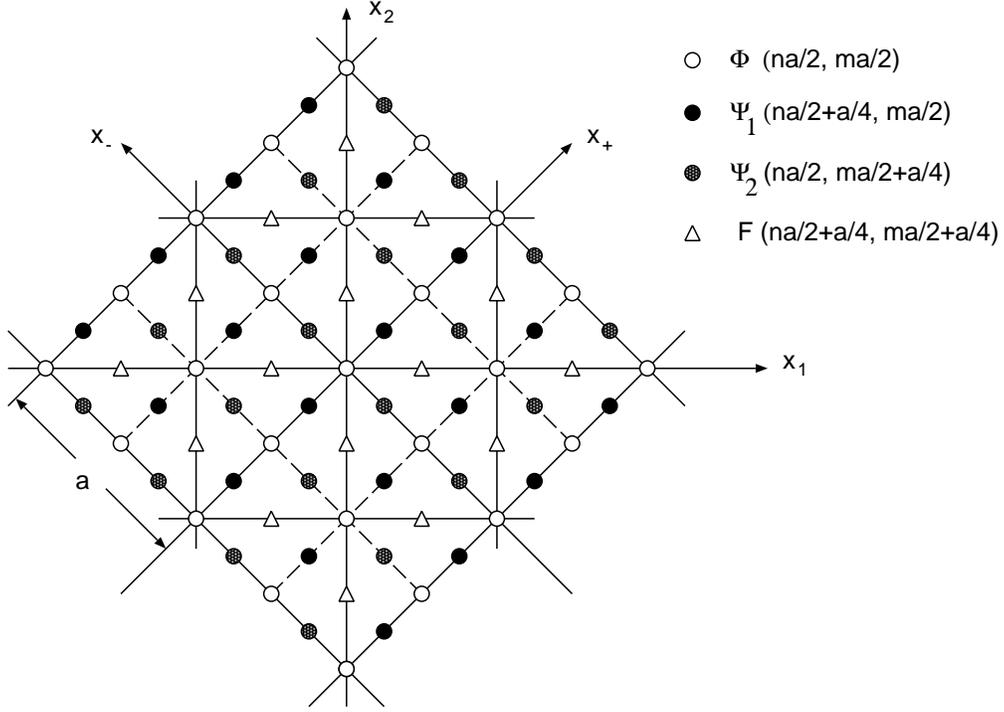}
 \caption{Fields defined on $a/4$ lattice sites.}
 \label{fig:field-scattered}
\end{figure}

In order to put this system on a lattice we proceed in the same way as in the $N=1$,$D=1$, treating
the two light cone variables independently. The lattice spacing is $\frac{a}{2}$ for both $x_+$ and $x_-$, but in order to introduce a half lattice difference operator we need to introduce sites on the multiples of $\frac{a}{4}$ as well.
A rather direct generalization of (\ref{spf}) is:
\beq
\Xi(x_+,x_-) = \left\{ \begin{array}{lc}  \Phi(x_+,x_-)~~~~~~~~&\textrm{for}~~~
 x_+,x_-=n a/2 ,\\\frac{1}{2}  e^{\frac{2i \pi x_+}{a}}
 \Psi_1(x_+,x_-) ~~~&\textrm{for}~~~x_+=(2n+1)a/4 ~~\textrm{and}~~ x_-=ma/2,\\
 \frac{1}{2}  e^{\frac{2i \pi x_-}{a}}
 \Psi_2(x_+,x_-) ~~~&\textrm{for}~~~x_-=(2n+1)a/4 ~~\textrm{and}~~ x_+=ma/2,\\
 -\frac{i}{4} e^{\frac{2i \pi (x_+ + x_-)}{a}} F(x_+,x_-)~~~~~~~~&\textrm{for}~~~
 x_+,x_-=(2n+1)a/4.
 \end{array} \right. \label{spf-2}
\eeq
See Fig.~\ref{fig:field-scattered} how the component fields are located
in the $a/4$ lattice sites.

As in the one dimensional case the supersymmetry transformation in one of the light cone directions can be defined as a finite difference over a half lattice spacing $\frac{a}{2}$ of the superfield $\Xi(x_+,x_-) $ :

\beq
\delta^{(1)}_+ \Xi(x_+,x_-) = \alpha_+ e^{\frac{2i\pi x_+}{a}} \left[ \Xi(x_+ +\frac{a}{4},x_-)-\Xi(x_+ -\frac{a}{4},x_-)\right] \label{susy1+},
\eeq
where $\alpha_+$ is the Grassmann odd parameters of the supersymmetry transformation.
The commutator of two $\delta^{(1)}_+$  transformations with parameters $\alpha_+$,$\beta_+$  gives a translation:
\beq
\left( \delta^{(1)}_{+\alpha}\delta^{(1)}_{+\beta}-\delta^{(1)}_{+\beta}\delta^{(1)}_{+\alpha}\right)
\Xi(x_+,x_-) = 2i\beta \alpha e^{\frac{4i\pi x_+}{a}} \left[ \Xi(x_+ +\frac{a}{2},x_-)-\Xi(x_+ -\frac{a}{2},x_-)\right] \label{trs}.
\eeq

In terms of the component fields the supersymmetry transformation (\ref{susy1+}) reads:
\bea
\delta^{(1)}_+\Phi(x_+,x_-) =& \frac{i\alpha_+}{2} \left[ \Psi_1(x_+ +\frac{a}{4},x_-) + \Psi_1(x_+ -\frac{a}{4},x_-)\right] \nonumber, \\
\delta^{(1)}_+\Psi_1(x_+,x_-) =& 2 \alpha_+ \left[ \Phi(x_+ +\frac{a}{4},x_-) - \Phi(x_+ -\frac{a}{4},x_-)\right], \nonumber \\
\delta^{(1)}_+\Psi_2(x_+,x_-) =& \frac{\alpha}{2} \left[ F(x_+ +\frac{a}{4},x_-) + F(x_+ -\frac{a}{4},x_-)\right], \nonumber \\
\delta^{(1)}_+F(x_+,x_-) =& 2i \alpha_+ \left[ \Psi_2(x_+ +\frac{a}{4},x_-) - \Psi_2(x_+ -\frac{a}{4},x_-)\right] \label{susy1+comp},
\eea
and the commutator (\ref{trs}) is, for any of the component fields:
\beq
\left( \delta^{(1)}_{+\alpha}\delta^{(1)}_{+\beta}-\delta^{(1)}_{+\beta}\delta^{(1)}_{+\alpha}\right)
\Phi_A(x_+,x_-) = 2i\alpha \beta  \left[ \Phi_A(x_+ +\frac{a}{2},x_-)-\Phi_A(x_+ -\frac{a}{2},x_-)\right], \label{trscomp}
\eeq
where $\Phi_A \equiv \{\Phi,\Psi_1,\Psi_2,F\}$.

The supersymmetry transformation along the other light cone direction $x_-$ requires some care.
Simply exchanging $x_+$ and $x_-$ in (\ref{susy1+}) would amount to exchanging $\Psi_1$ and $\Psi_2$, while we know from the continuum transformations (\ref{delta+}, \ref{delta-}) that $\delta^{(1)}_-$ is obtained from  $\delta^{(1)}_+$ by the exchanges $x_+ \leftrightarrow x_-$, $\psi_1 \leftrightarrow \psi_2$ and $F \leftrightarrow -F$\footnote{This comes from exchanging $\theta_+^{(1)}$ and $\theta_-^{(1)}$ in the superfield expansion (\ref{sup-N1})}.
The correct prescription for $\delta^{(1)}_-$ is to exchange in (\ref{susy1+}) $x_+$ and $x_-$ and replace at the same time the superfield  $\Xi(x_+,x_-)$ with its hermitian conjugate  $\Xi^\dag(x_+,x_-)$. With all component fields hermitian $\Xi^\dag$ is the same as $\Xi$ but for a change of sign of $\Psi_1$,$\Psi_2$ and $F$. An immaterial change of sign of the supersymmetry parameter $\alpha_-$ compensates for the sign change of $\Psi_1$ and $\Psi_2$ so that taking the hermitian conjugate of $\Xi$ amounts to changing the sign of $F$.

So the supersymmetry transformation $\delta^{(1)}_-$ can be written:
\beq
\delta^{(1)}_-\Xi^\dag(x_+,x_-) = -\alpha_-e^{\frac{2i\pi x_-}{a}} \left[ \Xi^\dag(x_+,x_- + \frac{a}{4}) -\Xi^\dag(x_+,x_- - \frac{a}{4})\right], \label{susy1-}
\eeq
or, taking the hermitian conjugate of both sides:
\beq
\delta^{(1)}_-\Xi(x_+,x_-) = - \left[ \Xi(x_+,x_- + \frac{a}{4}) -\Xi(x_+,x_- - \frac{a}{4})\right] \alpha_-e^{-\frac{2i\pi x_-}{a}}. \label{susy1-bis}
\eeq
A direct check, using (\ref{susy1-bis}), shows that the commutator of two supersymmetry transformations $\delta^{(1)}_-$ produces a translation, exactly as in the case of $\delta^{(1)}_+$ (\ref{trs}). The commutator of a $\delta^{(1)}_-$ and a $\delta^{(1)}_+$ transformation vanishes.
In terms of the component fields the supersymmetry transformation is, as expected, the same as (\ref{susy1+comp}) with the two light cone directions exchanged and an extra sign change of $F$:

\bea
\delta^{(1)}_-\Phi(x_+,x_-) =& \frac{i\alpha_-}{2} \left[ \Psi_2(x_+ ,x_- +\frac{a}{4}) + \Psi_2(x_+ ,x_- -\frac{a}{4})\right], \nonumber \\
\delta^{(1)}_-\Psi_2(x_+,x_-) =& 2 \alpha_- \left[ \Phi(x_+ ,x_- +\frac{a}{4}) - \Phi(x_+ ,x_- -\frac{a}{4})\right], \nonumber \\
\delta^{(1)}_-\Psi_1(x_+,x_-) =& -\frac{\alpha_-}{2} \left[ F(x_+ ,x_- +\frac{a}{4}) + F(x_+ ,x_- -\frac{a}{4})\right], \nonumber \\
\delta^{(1)}_-F(x_+,x_-) =& -2i \alpha_- \left[ \Psi_1(x_+ ,x_- +\frac{a}{4}) - \Psi_1(x_+ ,x_- -\frac{a}{4})\right] \label{susy1-comp}.
\eea

In conclusion, the supersymmetry transformations (\ref{susy1+comp}) and (\ref{susy1-comp}) provide a representation on the lattice of the $N=1$, $D=2$ susy algebra and reproduce the continuum transformations (\ref{delta+}) and (\ref{delta-}) in the continuum limit.
However a simple counting of the degrees of freedom shows that each field on the lattice corresponds to $4$ degrees of freedom in the continuum. This is due to the lattice spacing being $\frac{a}{2}$ while translational invariance is still associated to shifts of the original lattice spacing $a$. This means that for each fields on the lattice there are four translational invariant configurations (namely zero momentum configurations in the continuum). In fact a translational invariant configuration can be either constant or constant in absolute value but with alternating sign  along each light cone direction. So to each value of the momentum in the continuum along a light cone direction will correspond on the lattice two distinct values of the momentum defined on the lattice with spacing $\frac{a}{2}$.
This can be better seen in the momentum representation for the component fields defined by the Fourier transform and its inverse:
\bea
\Phi_A(p_+,p_-)=& \frac{1}{4} \sum_{x_+,x_-} e^{i(p_+x_+ + p_-x_-)} \Phi_A(x_+,x_-), \nonumber \\\Phi_A(x_+,x_-) =& \frac{a^2}{4\pi^2} \int_0^{\frac{4\pi}{a}} dp_+dp_- \Phi_A(p_+,p_-) e^{-i(p_+x_+ + p_-x_-)}, \label{Ftrans}
\eea
where as before $ \Phi_A\equiv \{\Phi,\Psi_1,\Psi_2,F\}$ and where in the sums  $x_+$ and $x_-$ are defined according to eq. (\ref{spf-2}).
In momentum representation the component fields are periodic (resp. antiperiodic) with period $\frac{4\pi}{a}$ whenever in coordinate
representation are defined on integer multiples of $\frac{a}{2}$ (resp. half-integer multiples). So we have:
\bea
\Phi(p_+,p_-)=&\Phi(p_+ +\frac{4\pi}{a},p_-)=\Phi(p_+,p_- +\frac{4\pi}{a}),\nonumber \\ \Psi_1(p_+,p_-)=&-\Psi_1(p_+ +\frac{4\pi}{a},p_-)=\Psi_1(p_+,p_- +\frac{4\pi}{a}), \nonumber \\\Psi_2(p_+,p_-)=&\Psi_2(p_+ +\frac{4\pi}{a},p_-)=-\Psi_2(p_+,p_- +\frac{4\pi}{a}), \nonumber \\ F(p_+,p_-)=&-F(p_+ +\frac{4\pi}{a},p_-)=-F(p_+,p_- +\frac{4\pi}{a}),\nonumber \\ \label{period2}
\eea

On the lattice the partial derivatives $\partial_{\pm}$ in the supersymmetry algebra (\ref{D=N=2-alg2}) are replaced by symmetric finite difference operators $\partial^{(s)}_{\pm}$ defined as in (\ref{symdiffop}), which corresponds in momentum representation to:
\beq
i\partial^{(s)}_{\pm} = 2 \sin\frac{ap_{\pm}}{2}. \label{sdo} \eeq

Requiring that $\partial^{(s)}_{\pm}$ obeys exact Leibnitz rule (and not some modified Leibnitz rule as in~\cite{DKKN}) amounts then to assume that the additive and conserved quantity are not the components $p_{\pm}$ of the momentum on the lattice but their sines $\sin\frac{ap_{\pm}}{2}$. The sine conservation law corresponds to introducing a non local ``star'' product in coordinate representation in place of the usual product as described in~\cite{D'Adda:2010pg}. In the continuum limit the conserved quantity $\frac{2}{a}\sin\frac{ap_{\pm}}{2}$ is the physical momentum. However, since $\sin\frac{ap_{\pm}}{2}$ is invariant under $p_{\pm} \rightarrow \frac{2\pi}{a} - p_{\pm}$ for each value of the momentum in the continuum there are four corresponding momentum configurations on the lattice. For instance zero momentum is represented on the lattice by: $\{p_+,p_-\}\equiv \{0,0\}, \{\frac{2\pi}{a},0\},\{0,\frac{2\pi}{a}\},\{\frac{2\pi}{a},\frac{2\pi}{a}\}$.
Correspondingly each lattice field $\Phi_A(p_+,p_-)$ corresponds in the continuum limit to four distinct degrees of freedom according to the following scheme:
\bea
&\Phi_A(p_+,p_-) \xrightarrow[ap_{\pm}\to 0]{}  \varphi^{00}_A(p_+,p_-),~&\Phi_A(\frac{2\pi}{a}-p_+,\frac{2\pi}{a}-p_-) \xrightarrow[ap_{\pm}\to 0]{}  \varphi^{11}_A(p_+,p_-),\nonumber \\ &\Phi_A(p_+,\frac{2\pi}{a}-p_-) \xrightarrow[ap_{\pm}\to 0]{}  \varphi^{01}_A(p_+,p_-),~&\Phi_A(\frac{2\pi}{a}-p_+,p_-) \xrightarrow[ap_{\pm}\to 0]{}  \varphi^{10}_A(p_+,p_-). \label{1to4}
\eea
The four lattice fields $\Phi_A$ represent then $16$ degrees of freedom in the continuum, the same number of degrees of freedom in the expansion of a $N=2$,$D=2$ superfield. We are going to show now that this is indeed what happens: the four lattice fields $\Phi_A$ transform as a representation of the $N=2$,$D=2$ supersymmetry algebra. Notice that in the case of fermion fields the four degrees of freedom in the continuum theory corresponding to, say, $\Psi_1$ are the result of the doubling phenomenon. The doubling problem is solved in this context by interpreting the doublers as members of the same $N=2$ supermultiplet.

First let us rewrite the supersymmetry transformations $\delta^{(1)}_\pm$ of eq.s (\ref{susy1+comp},\ref{susy1-comp}) in the momentum representation. They are:
\bea
Q_+^{(1)}\Phi(p_+,p_-)&=&i\cos \frac{ap_+}{4} \Psi_1(p_+,p_-),
\nonumber \\
Q_+^{(1)}\Psi_1(p_+,p_-)&=&-4i\sin \frac{ap_+}{4} \Phi(p_+,p_-),
\nonumber \\
Q_+^{(1)}\Psi_2(p_+,p_-)&=&\cos \frac{ap_+}{4} F(p_+,p_-),
\nonumber \\
Q_+^{(1)}F(p_+,p_-)&=& 4\sin \frac{ap_+}{4} \Psi_2(p_+,p_-),
\label{momdelta1+}
\eea
and:
\bea
Q_-^{(1)}\Phi(p_+,p_-)&=&i\cos \frac{ap_-}{4} \Psi_2(p_+,p_-),
\nonumber \\
Q_-^{(1)}\Psi_1(p_+,p_-)&=&-\cos \frac{ap_-}{4} F(p_+,p_-),
\nonumber \\
Q_-^{(1)}\Psi_2(p_+,p_-)&=&-4i\sin \frac{ap_-}{4} \Phi(p_+,p_-),
\nonumber \\
Q_-^{(1)}F(p_+,p_-)&=&-4\sin \frac{ap_-}{4} \Psi_1(p_+,p_-),
\label{momdelta1-}
\eea
where the supersymmetry parameter has been factored out by writing $\delta^{(1)}_\pm = \alpha_{\pm} Q_\pm^{(1)}$.

In order to obtain the second set of the $N=2$ supersymmetry transformations generated by $Q^{(2)}_\pm$ we make an ansatz based on the one dimensional case, where $Q_2$ transformations were obtained from $Q_1$ with the replacement (\ref{replacement}). In the present case the proposed ansatz is:
\bea
\Phi(p_+,p_-) \longrightarrow \Phi(p_+,p_-),&~~~~~~F(p_+,p_-)\longrightarrow -F(\frac{2\pi}{a}-p_+,\frac{2\pi}{a}-p_-), \nonumber \\
\Psi_1(p_+,p_-) \longrightarrow -i \Psi_1(\frac{2\pi}{a}-p_+,p_-),&~~~~~~\Psi_2(p_+,p_-) \longrightarrow -i \Psi_2(p_+,\frac{2\pi}{a}-p_-). \label{replacement2}
\eea
The replacement (\ref{replacement2}) leads for $Q^{(2)}_+$ and $Q^{(2)}_-$ to the following sets of transformations:
\bea
Q_+^{(2)}\Phi(p_+,p_-)&=&\cos \frac{ap_+}{4} \Psi_1(\frac{2\pi}{a}-p_+,p_-),
\nonumber \\
Q_+^{(2)}\Psi_1(p_+,p_-)&=&4\cos \frac{ap_+}{4} \Phi(\frac{2\pi}{a}-p_+,p_-),
\nonumber \\
Q_+^{(2)}\Psi_2(p_+,p_-)&=&-i\cos \frac{ap_+}{4} F(\frac{2\pi}{a}-p_+,p_-),
\nonumber \\
Q_+^{(2)}F(p_+,p_-)&=&4i\cos \frac{ap_+}{4} \Psi_2(\frac{2\pi}{a}-p_+,p_-),
\label{momdelta2+}
\eea
and
\bea
Q_-^{(2)}\Phi(p_+,p_-)&=&\cos \frac{ap_-}{4} \Psi_2(p_+,\frac{2\pi}{a}-p_-),
\nonumber \\
Q_-^{(2)}\Psi_1(p_+,p_-)&=&i\cos \frac{ap_-}{4} F(p_+,\frac{2\pi}{a}-p_-),
\nonumber \\
Q_-^{(2)}\Psi_2(p_+,p_-)&=&4\cos \frac{ap_-}{4} \Phi(p_+,\frac{2\pi}{a}-p_-),
\nonumber  \\
Q_-^{(2)}F(p_+,p_-)&=&-4i\cos \frac{ap_-}{4} \Psi_1(p_+,\frac{2\pi}{a}-p_-).
\label{momdelta2-}
\eea

A direct check shows that the supercharges defined by the above transformations
satisfy the $D=2$,$N=2$ supersymmetry algebra:
\beq
\{ Q_\pm^{(i)},Q_\pm^{(j)} \} =\delta^{ij}2\sin\frac{ap_\pm}{2}, ~~~
\{ Q_\pm^{(i)},Q_\mp^{(j)} \} =0~~
(i,j = 1,2).
\label{D=N=2-susy-alg}
\eeq

The continuum limit ($ap_\pm \rightarrow 0$) of the above supersymmetry  transformations can be done
keeping in mind that each lattice field splits into four component fields in the continuum according to the scheme of eq. (\ref{1to4}), namely:
\beq
\{\Phi,\Psi_1,\Psi_2,F\} \xrightarrow[ap_{\pm}\to 0]{} \{\varphi^{ij},\psi_1^{ij},\psi_2^{ij},f^{ij}\},
\label{1to4b}
\eeq
where the labels $i$ and $j$ take the values $\{0,1\}$ as in (\ref{1to4}). Each supersymmetry transformation, which on the lattice involves four fields will involve in the continuum all $16$ components labeled at the r.h.s. of (\ref{1to4b}). It is interesting to identify these with the components in a general superfield expansion. This identification requires a careful and lengthly comparison of the continuum limit of the supersymmetry transformations described above and the ones obtained in the continuum with the superfield formalism. The result is:
\bea
\Xi(x,\theta)&=& \varphi^{00}(x) + i\tup \psi_1^{00}(x) + i \tum \psi_2^{00}(x) + \tdp \psi_1^{10}(x) + \tdm \psi_2^{01}(x) +i \tup \tum f^{00}(x)\nonumber \\&-&  i \tdp \tdm f^{11}(x) +\tup \tdm f^{01}(x) -\tum \tdp f^{10}(x)-\tup \tdp \varphi^{10}(x) \nonumber \\ &-&  4i \tum \tdm \varphi^{01}(x) + 4i\tum \tdp \tdm \psi_1^{11}(x)-4i \tup \tdp \tdm \psi_2^{11}(x)  \nonumber\\ &+& 4\tup \tum \tdm \psi_1^{01}(x) - 4 \tup \tum \tdp \psi_2^{10}(x) + 16 \tup \tum \tdp \tdm \varphi^{11}(x). \label{csup}
\eea

Acting on (\ref{csup}) with the supersymmetry charges $Q_\pm^{(i)} = \frac{\partial}{\partial \theta_\pm^{(i)} } + i \theta_\pm^{(i)} \frac{\partial}{\partial x_\pm} $ one recovers the supersymmetry transformations in the continuum limit. They coincide with the ones obtained from (\ref{momdelta1+},\ref{momdelta1-},\ref{momdelta2+},\ref{momdelta2-}) by taking the continuum limit around the four zero momentum configuration. It appears from (\ref{csup}) that a lattice field generates, through its doublers, fields that have in general different dimensionality. For instance $\Phi(p_+,p_-)$ and $\Phi(\frac{2\pi}{a}-p_+,\frac{2\pi}{a}-p_-)$ correspond respectively to the first and last component in the superfield expansion. As the original lattice field was chosen to be dimensionless, different rescalings with different powers of the lattice constant $a$ will be needed to make contact with the fields in the continuum.
This feature was met already in the $N=2$, $D=1$ case~\cite{D'Adda:2010pg}, and will be discussed in detail further on.

\subsection{Chiral conditions}

In order to construct the two dimensional $N=2$ Wess-Zumino model on the lattice we need to reduce the general superfield, discussed in the previous subsection, to a chiral (or anti-chiral) superfield. This is normally done by imposing chiral conditions, and this requires to define  the lattice counterpart of the super derivative.
The definition of super derivative proceeds along the lines followed in section \ref{section3} for the one dimensional case. In the continuum superfield formulation the super derivative is obtained from  the corresponding supercharge by changing the sign of $\frac{\partial}{\partial x^\mu}$. On the lattice, as explained in section \ref{section3}, this corresponds to the following replacements at the r.h.s. of the supersymmetry transformations (\ref{momdelta1+},\ref{momdelta1-},\ref{momdelta2+},\ref{momdelta2-}):
\beq
\cos\frac{ap_\pm}{4}~\rightarrow~\cos\frac{ap_\pm}{2}\cos\frac{ap_\pm}{4},~~~
\sin\frac{ap_\pm}{4}~\rightarrow~-\cos\frac{ap_\pm}{2}\sin\frac{ap_\pm}{4}.
\label{change-to-susyder-2d}
\eeq
Let us denote by $D_\pm^{(i)}$ with $i=\{1,2\}$ the super derivatives. The action of $D_\pm^{(i)}$ on the lattice fields can then be obtained with the substitution (\ref{change-to-susyder-2d}) in (\ref{momdelta1+},\ref{momdelta1-},\ref{momdelta2+},\ref{momdelta2-}). In order to impose chiral conditions however it is more convenient to introduce directly the chiral super derivatives defined as:
\beq
D_\pm^{(\pm)} = \frac{1}{2} (D_\pm^{(1)} \pm i D_\pm^{(2)}).
\label{def-chi-scov-der}
\eeq
The action of $D_\pm^{(\pm)}$ on the lattice fields is:

\bea
D_+^{(\pm)}\Phi(p_+,p_-)&=&\frac{i}{2}\cos \frac{ap_+}{2}\cos \frac{ap_+}{4}
\left[\Psi_1(p_+,p_-) \pm \Psi_1(\frac{2\pi}{a}-p_+,p_-) \right],
\nonumber \\
D_+^{(\pm)}\Psi_1(p_+,p_-)&=&2i\cos \frac{ap_+}{2}\left[ \sin \frac{ap_+}{4}
\Phi(p_+,p_-) \pm \cos \frac{ap_+}{4}\Phi(\frac{2\pi}{a}-p_+,p_-) \right],
\nonumber \\
D_+^{(\pm)}\Psi_2(p_+,p_-)&=&\frac{1}{2}\cos \frac{ap_+}{2}\cos \frac{ap_+}{4}\left[
F(p_+,p_-) \pm F(\frac{2\pi}{a}-p_+,p_-)\right],
\nonumber \\
D_+^{(\pm)}F(p_+,p_-)&=& -2\cos \frac{ap_+}{2}\left[ \sin \frac{ap_+}{4}
\Psi_2(p_+,p_-)\pm \cos \frac{ap_+}{4}
\Psi_2(\frac{2\pi}{a}-p_+,p_-)\right] ,
\nonumber \\
D_-^{(\pm)}\Phi(p_+,p_-)&=&\frac{i}{2}\cos \frac{ap_-}{2}\cos \frac{ap_-}{4}
\left[ \Psi_2(p_+,p_-) \pm \Psi_2(p_+,\frac{2\pi}{a}-p_-) \right],
\nonumber \\
D_-^{(\pm)}\Psi_1(p_+,p_-)&=&-\frac{1}{2}\cos \frac{ap_-}{2}\cos \frac{ap_-}{4}
\left[ F(p_+,p_-) \pm F(p_+,\frac{2\pi}{a}-p_-)\right],
\nonumber \\
D_-^{(\pm)}\Psi_2(p_+,p_-)&=&2i\cos \frac{ap_-}{2}\left[ \sin \frac{ap_-}{4}
\Phi(p_+,p_-)\pm \cos \frac{ap_-}{4} \Phi(p_+,\frac{2\pi}{a}-p_-)\right],
\nonumber \\
D_-^{(\pm)}F(p_+,p_-)&=&2\cos \frac{ap_-}{2}\left[ \sin \frac{ap_-}{4}
\Psi_1(p_+,p_-)\pm \cos \frac{ap_-}{4}\Psi_1(p_+,\frac{2\pi}{a}-p_-)\right],
\label{2chiralder}
\eea
A direct check shows that $D_\pm^{(\pm)}$ anticommute with all supersymmetry charges and satisfy
the following algebra:
\beq
\{D^{(+)}_+,D^{(-)}_+\} = -2 \cos^2\frac{ap_+}{2}\sin\frac{ap_+}{2},~~~~~\{D^{(+)}_-,D^{(-)}_-\} = -2 \cos^2\frac{ap_-}{2}\sin\frac{ap_-}{2},
\label{Dalg}
\eeq
with all other anticommutators vanishing.
Chiral conditions are obtained by imposing
\beq
D_+^{(-)} \Phi_A(p_+,p_-) = 0, ~~~~~D_-^{(-)} \Phi_A(p_+,p_-) = 0,\label{chiral}
\eeq
for $\Phi_A =\{ \Phi, \Psi_1, \Psi_2, F\}$
and can be read directly at the r.h.s. of (\ref{2chiralder}) as they require the vanishing of the square brackets when the minus sign is chosen. Anti-chiral conditions are similarly obtained by exchanging the minus with the plus sign.
Both chiral and anti-chiral conditions are better expressed by introducing a set of rescaled fields defined as:
\bea
\uphi(p_+,p_-) &=&
\frac{\Phi(p_+,p_-)}{\cos\frac{ap_+}{4}\cos\frac{ap_-}{4}}, ~~~\underline{F}(p_+,p_-) = F(p_+,p_-),
\nonumber \\
\ups(p_+,p_-)&=& \frac{\Psi_1(p_+,p_-)}{\cos\frac{ap_-}{4}}, ~~~
\upS(p_+,p_-)\equiv \frac{\Psi_2(p_+,p_-)}{\cos\frac{ap_+}{4}}.
\label{def-chi-fields-2d}
\eea
In fact in terms of these fields chiral (resp. anti-chiral) conditions are simply equivalent to symmetrization
(resp anti-symmetrization) with respect to symmetry operations $ p_+ \rightarrow \frac{2\pi}{a}-p_+$ and $p_- \rightarrow \frac{2\pi}{a}-p_-$. So chiral conditions just give:
\beq
\uphi_A(p_+,p_-) = \uphi_A(\frac{2\pi}{a}-p_+,p_-)=\uphi_A(p_+,\frac{2\pi}{a}-p_-)=\uphi_A(\frac{2\pi}{a}-p_+,\frac{2\pi}{a}-p_-),
\label{chiral2}
\eeq
with $\uphi_A \equiv \{\uphi,\ups,\upS,\underline{F}=F\}$.
Anti-chiral conditions are:
\beq
\uphi_A(p_+,p_-) = -\uphi_A(\frac{2\pi}{a}-p_+,p_-)=-\uphi_A(p_+,\frac{2\pi}{a}-p_-)=\uphi_A(\frac{2\pi}{a}-p_+,\frac{2\pi}{a}-p_-).
\label{antichiral}
\eeq

The chiral conditions written in (\ref{chiral2}) reproduce in the continuum limit the usual chiral condition for a $D=2$, $N=2$ superfield. For example, consider
$$ \uphi(p_+,p_-) = \uphi(\frac{2\pi}{a}-p_+,\frac{2\pi}{a}-p_-). $$
Going back to the non underlined fields and taking the continuum limit we find (with the notations of eq. (\ref{csup}) ):
$$ \varphi^{11}(x) = -\frac{1}{16} \frac{\partial^2}{\partial x_+\partial x_-} \varphi^{00}(x), $$
which is the usual relation between the first and last component of a chiral $D=2$,$N=2$ superfield.
On the other hand, in terms of the underlined lattice fields, the chiral conditions simply state the identification (possibly modulo a sign)  of the four different doublers for each field. This identification
has been made possible by having halved the lattice size and hence doubled the Brillouin zone.

\begin{figure}
 \hfil\includegraphics[height=0.6\linewidth]{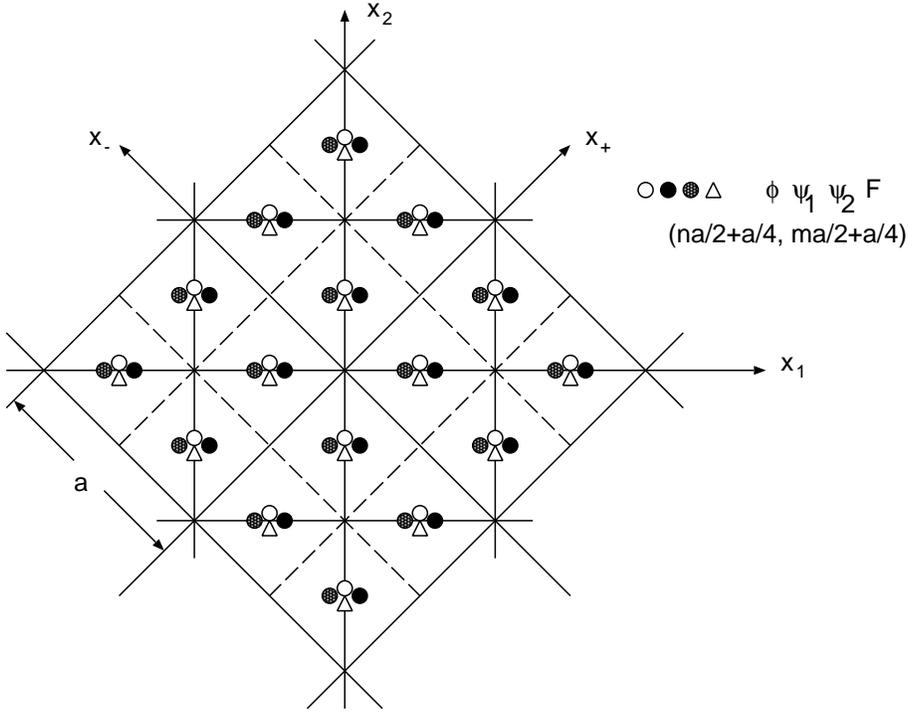}
 \caption{All rescaled fields gather on ($a/4,a/4$) lattice sites.}
 \label{fig:field-collected}
\end{figure}

The underlined fields $\uphi_A$ have different periodicity in $p_\pm$ with respect to $\Phi_A$. In fact, according to (\ref{period2}) and (\ref{def-chi-fields-2d}) they are all anti-periodic in both $p_+$ and $p_-$.
In coordinate representation this means that they are all defined on sites $(n+\frac{1}{2})\frac{a}{2}$.
See Fig.~\ref{fig:field-collected} to see how the semi locally scattered fields
shown in Fig.~\ref{fig:field-scattered} are shifted to the newly rescaled
fields after the chiral conditions are imposed.
It should be noticed however that the representation on the lattice of the $D=2$, $N=2$ susy algebra in terms of chiral superfields given below is consistent with either periodic or antiperiodic conditions in the momenta, so that both can be consistently chosen. For a detailed discussion of periodicity, symmetry and hermiticity properties of the chiral and antichiral fields see Appendix A.  In the next section chiral and antichiral representations with periodic fields will be used to construct the supersymmetric action on the lattice. This would correspond in Fig.~\ref{fig:field-scattered} to have the origin shifted of $\frac{a}{4}$ in each direction and all fields sitting on sites $(n\frac{a}{2}+\frac{a}{4},m\frac{a}{2}+\frac{a}{4})$.

In the continuum limit ( $\sin\frac{ap_\pm}{2}\to 0$) $\Phi_A$ and $\uphi_A$ coincide if the limit is taken as $ap_\pm \to 0$ since $\cos\frac{ap_\pm}{4}$ in (\ref{def-chi-fields-2d}) is one in that limit. But if the limit is taken with either $p_+$ or $p_-$ going to $\frac{2\pi}{a}$ a derivative is generated by the cosine factor.
For instance, using for the underlined fields the same notations used in (\ref{1to4}) we have:
$\psi_1^{01} = \frac{1}{4} ap_- \underline{\psi_1^{01}}$.
However as a result of the chirality  conditions (\ref{chiral2}) the degrees of freedom describing fluctuations around the $p_\pm=0$ vacuum are the only independent ones, so the supersymmetry transformations for  chiral (and anti-chiral) superfields can be conveniently described in terms of $\uphi_A$.
In order to do that it is convenient to introduce chiral supercharges:

\beq
Q_\pm^{(\pm)} =\frac{1}{2} \left( Q_\pm^{(1)} \pm i Q_\pm^{(2)} \right) \label{chsup},
\eeq
which satisfy the algebra:
\beq
\left\{Q_\pm^{(+)},Q_\pm^{(-)}\right\} = 2 \sin\frac{ap_\pm}{2},  \label{chiralalgebra}
\eeq
with all other anticommutators vanishing.
The supersymmetry transformations for chiral and anti-chiral superfields are then given respectively in table
2 and 3. In table 3 the components of the anti-chiral superfield are denoted, to distinguish them from the chiral ones, with an over line.

\begin{table}
\hfil
$
\def\arraystretch{1.2}
\begin{array}{|l|c|c|c|c|} \hline
   &\displaystyle  Q_+^{(+)} & \displaystyle Q_+^{(-)}
&\displaystyle Q_-^{(+)} &\displaystyle Q_-^{(-)} \\ \hline
\hline
\uphi(p) & i\ups(p) & 0& i\upS(p) & 0 \\
\ups(p) & 0 & -2i\sin \frac{ap_+}{2} \uphi(p) & -F(p) & 0 \\
\upS(p) & F(p) &  0 & 0 & -2i\sin \frac{ap_-}{2}\uphi(p) \\
F(p) & 0 & 2\sin \frac{ap_+}{2} \upS(p) & 0
&-2\sin \frac{ap_-}{2} \ups(p)\\ \hline
\end{array}
$
\caption{Chiral $D=N=2$ supersymmetry transformation }
\end{table}

\begin{table}
\hfil
$
\def\arraystretch{1.2}
\begin{array}{|l|c|c|c|c|} \hline
   &\displaystyle  Q_+^{(+)} & \displaystyle Q_+^{(-)}
&\displaystyle Q_-^{(+)} &\displaystyle Q_-^{(-)} \\ \hline
\hline
\bar{\uphi}(p) & 0 & i\bar{\ups}(p) & 0  & i\bar{\upS}(p)  \\
\bar{\ups}(p) & -2i\sin \frac{ap_+}{2} \bar{\uphi}(p) & 0 & 0 & -\bar{F}(p)  \\
\bar{\upS}(p) & 0& \bar{F}(p) &  -2i\sin \frac{ap_-}{2}\bar{\uphi}(p)& 0 \\
\bar{F}(p) & 2\sin \frac{ap_+}{2} \bar{\upS}(p) & 0
&-2\sin \frac{ap_-}{2} \bar{\ups}(p) & 0 \\ \hline
\end{array}
$
\caption{anti-chiral $D=N=2$ supersymmetry transformation }
\end{table}


\section{ N=2 Wess-Zumino model in two dimensions}
\label{section4}
\setcounter{equation}{0}

In the previous section the chiral and anti-chiral representations of the $N=2$, $D=2$ supersymmetry algebra given in tables 2 and 3 have been derived by applying chiral conditions on the reducible representation corresponding to the full superfield expansion.
Chiral and anti-chiral representations are irreducible representations, and so they are the building blocks of any supersymmetric action. The result shown in tables 2 and 3 can then be regarded as valid in itself, independently on how it has been derived.

The lattice representations of table 2 and 3 can be  obtained from the corresponding representations in the continuum by simply replacing the derivatives $p_\pm$ with the sine variables $\sin \frac{ap_\pm}{2}$. This implies  a lattice spacing $\frac{a}{2}$  (hence a periodicity in momentum space of $\frac{4\pi}{a}$) and consequently a doubling of the degrees of freedom. The doubling can be removed by assuming that all fields are symmetric (or anti-symmetric) in momentum representation under the symmetry operations $p_\pm \to \frac{2\pi}{a}-p_\pm$. This leads to the conditions (\ref{chiral2}) or (\ref{antichiral}) that in the previous section were obtained by imposing chiral conditions on the full superfield.

The supersymmetry transformations listed in tables 2 and 3 are consistent with all fields being periodic or all anti-periodic with period $\frac{4\pi}{a}$. They are also consistent with fields being symmetric or antisymmetric with respect to $p_\pm \to \frac{2\pi}{a}-p_\pm$.
However if we assume that anti-chiral fields are hermitian conjugate of the chiral ones, so that:
\beq
\uphi_A^{\dagger}(p) = \bar{\uphi}_A(-p),
\label{herm}
\eeq
then the periodicity with period $\frac{4\pi}{a}$ in the momenta implies that chiral and anti-chiral fields are both symmetric or both antisymmetric with respect to $p_\pm \to \frac{2\pi}{a}-p_\pm$. On the other hand anti periodicity in the momenta requires chiral and anti-chiral fields to have opposite symmetry properties with respect to $p_\pm \to \frac{2\pi}{a}-p_\pm$ as in (\ref{chiral2}) and (\ref{antichiral}).
All these choices  are consistent with the supersymmetry transformations. However writing the action imposes some restrictions. The Lagrangian density has to be periodic with period $\frac{4\pi}{a}$  in each integrated momentum. An antiperiodic density would lead to a vanishing action when integrated over a  $\frac{8\pi}{a}$ interval. Since each integrated momentum is associated to just one field the anti-periodic option for the fields should be discarded\footnote{A trigonometric function with argument $\frac{ap}{4}$ could be introduced to compensate for the anti-periodicity of the field itself, but that could always be absorbed in the field definition.}.
Let us now consider the symmetry under $p_\pm \to \frac{2\pi}{a}-p_\pm$. Let $L$ be the Lagrangian density and $p$ an integration variable, namely the $+$ or $-$ component of a field's momentum. If we use $dp$ as integration measure the Lagrangian density should be symmetric (and not antisymmetric) under $p \to \frac{2\pi}{a}-p$. In fact, if $L(\cdots,\frac{2\pi}{a}-p,\cdots)=L(\cdots,p,\cdots)$ we have:
\beq
\int_{-\frac{\pi}{a}}^{\frac{3\pi}{a}} dp L(\cdots,p,\cdots) = 2 \int_{-\frac{\pi}{a}}^{\frac{\pi}{a}} dp L(\cdots,p,\cdots),
\label{twice}
\eeq
whereas the l.h.s. of (\ref{twice}) vanishes if $L(\cdots,\frac{2\pi}{a}-p,\cdots)=-L(\cdots,p,\cdots)$ .
If all the $p$ dependence in $L$ is contained in the corresponding fields and in the delta function expressing the sine momentum conservation, then all fields should be symmetric under $p_\pm \to \frac{2\pi}{a}-p_\pm$.
 However if we use $\hat{p}= \frac{2}{a}\sin\frac{ap}{2}$ as integration variable and $d\hat{p}=\cos\frac{ap}{2}dp$ as integration volume the extra antisymmetric cosine factor will require an antisymmetric Lagrangian density to re-establish the symmetry. In this case all fields should be antisymmetric under  $p_\pm \to \frac{2\pi}{a}-p_\pm$.

 The two choices are  equivalent: in the former case the cosine factor is included in the field definition, in the latter case in the integration volume and multiplication by a factor $\cos\frac{ap_+}{2} \cos\frac{ap_-}{2}$  changes the symmetry property of the field under $p_\pm \to \frac{2\pi}{a}-p_\pm$ while not affecting its continuum limit\footnote{Notice however that a symmetric field is unconstrained, while an antisymmetric one vanishes at $p_\pm = \pm
\frac{\pi}{a}$.}.

\subsection{The kinetic term}

In our lattice formulation the quadratic terms in the action enjoy a unique property, as already discussed in the one dimensional case of ref.~\cite{D'Adda:2010pg}. The sine momentum conservation
\beq
\sin\frac{ap_\pm}{2} + \sin\frac{aq_\pm}{2} = 0,   \label{2pt}
\eeq
has two separate solutions, namely:
\beq
p_\pm + q_\pm = 0,~~~~~~~~~~\textrm{and}~~~~~~~~~~~~~p_\pm - q_\pm = \frac{2\pi}{a},~~~~~~~~~~~~~~~~~~ (\textrm{mod}\frac{4\pi}{a}),   \label{2ptc}
\eeq
so that the delta functions expressing the sine conservation laws can be replaced, preserving all the symmetries of the theory, by delta functions with the argument linear in the momenta, namely by:
\beq
\delta^{(s)}(p+q) \equiv \prod_{i=\pm} \frac{1}{2}\left[ \delta(p_i+q_i)+\delta(p_i-q_i+\frac{2\pi}{a}) \right],~~~~~~~~(\textrm{mod}\frac{4\pi}{a}).
\label{lcons}
\eeq

The kinetic term of the supersymmetric Wess-Zumino action can then be written on the lattice using the same construction  as in the continuum:
\bea
S_K &=&\int_{-\frac{\pi}{a}}^{\frac{3\pi}{a}} dp_+ dp_- dq_+ dq_-  \delta^{(s)}(p+q)  Q_+^{(-)}Q_-^{(-)}Q_+^{(+)}Q_-^{(+)} \{\bar{\uphi}(p)\uphi(q)\}
\nonumber \\
&=&\int_{-\frac{\pi}{a}}^{\frac{3\pi}{a}} dp_+ dp_- dq_+ dq_-  \delta^{(s)}(p+q)
 \left[-4\bar{\uphi}(p) \sin\frac{aq_+}{2}\sin\frac{aq_-}{2}\uphi(q) -\bar{F}(p) F(q)\right.\nonumber \\
 &&\left.+2\bar{\upS}(p)\sin\frac{aq_+}{2} \upS(q) +
2\bar{\ups}(p)\sin\frac{aq_-}{2}\ups(q)\right].
\label{D=N=2-WZ-action}
\eea
The invariance of the action $S_K$ under all the supersymmetry transformations generated by $Q_{\pm}^{(\pm)}$ is assured, as in the continuum, by the algebra of the $Q$'s and by the momentum conservation (\ref{lcons}), which is equivalent to (\ref{2pt}).

We assume that all fields satisfy the symmetry condition (\ref{chiral2}), so that in each variable the contribution of the integration in the intervals $(-\frac{\pi}{a},\frac{\pi}{a})$ and $(\frac{\pi}{a},\frac{3\pi}{a})$ coincide and we get:

\bea
S_K &=& 4 \int_{-\frac{\pi}{a}}^{\frac{\pi}{a}} dp_+ dp_- dq_+ dq_-  \delta(p_+ +q_+)\delta(p_- +q_-)
 \left[-4\bar{\uphi}(p) \sin\frac{aq_+}{2}\sin\frac{aq_-}{2}\uphi(q) \right.\nonumber \\
 &&\left. -\bar{F}(p) F(q)+2\bar{\upS}(p)\sin\frac{aq_+}{2} \upS(q) +
2\bar{\ups}(p)\sin\frac{aq_-}{2}\ups(q)\right].
\label{D=N=2-WZ-action2}
\eea

The advantage of using directly the momentum conservation (\ref{lcons}), rather than the sine conservation, is that the action (\ref{D=N=2-WZ-action}) can be formulated in coordinate representation by a simple Fourier transform. Using (\ref{chiral2}) we find:
\bea
S_K &=& 4 \pi^2 \sum_{x_\pm = \frac{n_\pm a}{2}} \left[ \bar{\uphi}(x)\Delta_+\Delta_-\uphi(x) -\bar{F}(x)F(x) -i \bar{\upS}(x)\Delta_+\upS(x) \right.
\nonumber \\ &&\left. -i \bar{\ups}(x) \Delta_- \ups(x) \right], \label{kinx}
\eea
where
\bea
\Delta_+ \uphi_A(x) &=& \uphi_A(x_+ +\frac{a}{2},x_-)-\uphi_A(x_+ -\frac{a}{2},x_-), \nonumber \\ \Delta_- \uphi_A(x)& =& \uphi_A(x_+,x_-+\frac{a}{2})-\uphi_A(x_+,x_- -\frac{a}{2}). \label{Delta}
\eea
The fields in action (\ref{kinx}) must satisfy the ``chiral'' conditions (\ref{chiral2}), which in coordinate representation are:
\beq
\uphi_A(x_+,x_-) = (-1)^{n_+} \uphi_A(-x_+,x_-)=(-1)^{n_-} \uphi_A(x_+,-x_-)=(-1)^{n_+ + n_-} \uphi_A(-x_+,-x_-), \label{xminusx}
\eeq
with $x_{\pm}=\frac{a n_\pm }{2}$.
Through eq.s (\ref{xminusx}) the degrees of freedom in four different points, obtained by reflection with respect to the $x_+=0$ and $x_-=0$ axis,
are identified and in spite of the $a/2$ lattice spacing the density of independent degrees of freedom is the same as in a lattice with spacing $a$.
The lattice is not smoothly mapped into the coordinate space in the continuum limit as eq.s (\ref{xminusx}) have no continuum limit. To make contact with continuum space-time one has to separate long wavelength modes ( around $p_\pm=0$) from the short wavelength ones (around $p_\pm=\frac{2\pi}{a}$) using (\ref{chiral2}). In other words one has to consider the action (\ref{D=N=2-WZ-action2}), where all fields are defined in the momentum region $
-\frac{\pi}{a}\leq p_\pm \leq \frac{\pi}{a}$. The Fourier transform over a $\frac{2\pi}{a}$ interval will produce fields
in coordinate representation defined on a lattice with spacing $a$, which can be
identified with the continuum space-time in the continuum limit.  With this lattice
we will show in Appendix that the translational invariance, which is broken
by the sine conservation law, is recoverd in the continuum limit.
However the symmetric finite difference operators $\sin\frac{ap_\pm}{2}$
are not periodic with period $\frac{2\pi}{a}$ , so the resulting action
in coordinate representation will be highly non-local as in the SLAC derivative case.

The kinetic term (\ref{kinx}) instead is local on the $a/2$ lattice,
modulo the $x \to -x$ symmetry (\ref{xminusx}), which also breaks the
translational invariance.

All fields appearing in (\ref{kinx}) are dimensionless. When the continuum limit is taken some powers of $a$ appear and fields need to scale with $a$ for the limit to be smooth. More precisely we have  that powers of $a$ are generated by $ \sum_{n_\pm} \to \frac{1}{a^2} \int dx_+dx_-$ and by $\Delta_\pm \to a \frac{\partial}{\partial x_\pm }$. In order for these powers of $a$ to be absorbed fields have to rescale with $a$, the rescaled fields thus acquiring the standard dimensions fields. So we have:
\beq
\uphi(x) \to \varphi(x),~~~~\frac{1}{\sqrt{a}} \underline{\Psi_i}(x) \to  \psi_i(x),
~~~~\frac{1}{a} F(x) \to  f(x)  \label{rescaling}
\eeq
where the fields at the right sides are the rescaled dimensional fields, to be identified with the ones in the continuum limit.

\subsection{Mass term and interaction terms}

As in the continuum formulation mass and interaction term have a different structure with respect to the kinetic term, as they made by the sum of two terms, one involving only chiral fields, the other only anti-chiral fields.
The general structure is:
\bea
S_n &=&\int \prod_{j=1}^n  d^2p_j V_n(p)
Q_+^{(+)} Q_-^{(+)}\{ \uphi(p_1)\uphi(p_2) \cdots \uphi(p_n) \} + \textrm{h.c.}
\label{int1} \\
&=& \int \prod_{j=1}^n  d^2p_j V_n(p) n \left[i F(p_1)\prod_{j=2}^n \uphi(p_j) +(n-1) \upS(p_1)\ups(p_2) \prod_{j=3}^n \uphi(p_j) \right] +\textrm{h.c.},
\nonumber
\eea
where $V_n(p)$ is an integration volume to be discussed shortly which includes the delta function of momentum conservation and is symmetric under permutations of the $n$ momenta.

In the case of the mass term we set $n=2$ in (\ref{int1}). Moreover, being a quadratic interaction, we can simply choose, as in the kinetic term, the following integration volume:
\beq
V_2(p) = m a^2 \delta^{(s)}(p_1+p_2) =  m a^2 \prod_{i=\pm} \frac{1}{2}\left[ \delta(p_{1i}+p_{2i}) + \delta(p_{1i}-p_{2i}+\frac{2\pi}{a}) \right] ~~~~\textrm{mod.}\frac{4\pi}{a}, \label{massvol}
\eeq
where $m$ is an dimensionless mass parameter and the factor $a^2$ is introduced for dimensional reasons so that $S_2$ is dimensionless.
So the mass term in momentum representation is:

\beq
S_2 = m a^2  \int \prod_{j=1}^2 d^2p_j~ \delta^{(s)}(p_1+p_2) \left[ i F(p_1)\uphi(p_2)+ \upS(p_1)\ups(p_2) \right], \label{mass1}
\eeq
where as usual the ``chiral'' conditions (\ref{chiral2}) hold. As in the kinetic term the action is local in coordinate representation, but with the non-local $x \to -x$ symmetry (\ref{xminusx}):

\beq
S_2 = 4 m \pi^2 \sum_{x_\pm = \frac{a n_\pm}{2}}  \left[ i F(x)\uphi(x) + \upS(x)\ups(x) \right].
\label{mass2}
\eeq

To make contact with the continuum theory the lattice fields must be rescaled as in (\ref{rescaling}) and the resulting $a$ factors are absorbed in the parameter $m$ leading to the dimensional mass parameter $M = \frac{m}{a}$.

In the interaction terms ($n \geq 3$) the integration volume $V_n(p)$ must contain the delta function expressing the sine momentum conservation. We define then:
\beq
V_n(p) = a^{2n}~ g_n~ G_n(p)~\delta^{(2)}\left(\sin\frac{ap_1}{2}+\sin\frac{ap_2}{2}+\cdots+\sin\frac{ap_n}{2}\right),
\label{sdelta}
\eeq
where $g_n$ is a dimensionless coupling constant and $G_n(p)$ can be in principle any function periodic in each momentum component with period $\frac{4\pi}{a}$, symmetric under permutations of the momenta and invariant under the substitution $ p_{i\pm} \to \frac{2\pi}{a} - p_{i\pm}$. We choose $G_n(p)$  to be $1$  in the continuum limit, namely when all momentum components  $p_{i\pm}$ are either zero or $\frac{2\pi}{a}$. We also choose $G_n(p)$ to vanish whenever a momentum component $p_{i\pm}$ is equal to $\pm\frac{\pi}{a}$ to avoid the integration volume to blow up as an effect of the sines in the delta function. The minimal way to satisfy all these requirements is to choose:
\beq
G_n(p) = C_n(p),
\label{GC}
\eeq
where
\beq
C_n(p) \equiv \prod_{j=1}^n | \cos\frac{ap_{j+}}{2} \cos\frac{ap_{j-}}{2} |.
\label{C}
\eeq
Clearly any positive power of $C_n(p)$ ( for instance $G_n(p) = C_n^2(p)$) would equally satisfy all symmetry requirement.

As an example we give explicitly the cubic interaction term. Setting $n=3$ in (\ref{int1}) and using (\ref{sdelta}) and (\ref{GC}) we have:
\bea
S_3 &=&3 a^6 g_3\int \prod_{j=1}^3  d^2p_j |\cos\frac{ap_{j+}}{2}\cos\frac{ap_{j-}}{2}| ~\delta^{(2)}\left(\sin\frac{ap_1}{2}+\sin\frac{ap_2}{2}+\sin\frac{ap_3}{2}\right) \nonumber \\
& &\left[i F(p_1) \uphi(p_2)\uphi(p_3) +2 \upS(p_1)\ups(p_2) \uphi(p_3) \right] +\textrm{h.c.}. \label{S3}
\eea
The continuum limit of (\ref{S3}) is obtained by taking $a \to 0$, namely $|a p_j| \ll 1$. In principle we should also consider the regions where one or more components of the momenta are close to $\frac{2\pi}{a}$, but thanks to (\ref{chiral2}) these regions are just copies of the $p_j=0$ region and they only give an overall multiplicative factor. All fields in (\ref{S3}) are dimensionless, so a rescaling is needed to make contact with the continuum fields:
\beq
a^2 \uphi(p) \to \varphi(p),~~~~a^{\frac{3}{2}} \underline{\Psi_i}(p) \to  \psi_i(p),
~~~~a F(p) \to  f(p),  \label{rescalingp}
\eeq
which differ from (\ref{rescaling}) because of the different dimensionality in the continuum of fields in coordinate and momentum representation.
Neglecting overall numerical factors the continuum limit of (\ref{S3}) is then:
\beq
S_3 \to g^{(c)}_3 \int \prod_{j=1}^3  d^2p_j \delta^{(2)}\left(p_1+p_2+p_3\right)\left[i f(p_1) \varphi(p_2)\varphi(p_3) +2 \psi_2(p_1)\psi_1(p_2) \varphi(p_3) \right] +\textrm{h.c.}, \label{S3cont}
\eeq
where $g^{(c)}_3 = \frac{g_3}{a^3}$ is the dimensional coupling constant of the continuum theory.

As in the one dimensional case of ref.~\cite{D'Adda:2010pg}
the coordinate representation of the interaction terms involves non-locality,
due to the sine momentum conservation.
This can be formalized by introducing, as in~\cite{D'Adda:2010pg}, a new non-local product of fields, the star product. With respect to this product the
symmetric finite difference operator satisfies the Leibnitz rule, leading in momentum space to the sine momentum conservation.
This will be explained in detail in the next section.

Another affect of the sine conservation law is that it breaks
the translational invariance at finite lattice spacing.
In the continuum limit, however, the translational invariance
is recovered (see Appendix~\ref{sec:trans-inv}).

\section{ Star product in two dimensions}
\label{section6}
\setcounter{equation}{0}

 With ordinary momentum conservation
the convolution of a product of two field $F \cdot G$ is defined
as a result of additive momentum $p=p_1+p_2$ under the product:
\beq
(F\cdot G)(p) = \left(\frac{a}{2\pi}\right)^2\int d^2 p_1 d^2 p_2 F(p_1) G(p_2)
\delta^{(2)}(p-p_1-p_2). \label{dotprod}
\eeq
 In coordinate space this amounts to the ordinary local product:
 \beq
  (F \cdot G)(x) = F(x) G(x).
 \label{dotcoord}
 \eeq
On the lattice the standard momentum conservation is replaced by the
lattice (sine) momentum conservation, which means that
$\hat{p}_\pm= \frac{2}{a} \sin\frac{a p_\pm}{2}$
is the additive quantity when taking the product of two fields.
 This amounts to changing the definition of the ``dot'' product  to that of a ``star'' product defined in momentum space as:
 \beq
 (F * G)(p)
= \left(\frac{a}{2\pi}\right)^2\int d^2 \hat{p}_1 d^2 \hat{p}_2 F(p_1) G(p_2)
\delta^{(2)}(\hat{p}-\hat{p}_1-\hat{p}_2)   \label{starprod}
\eeq

As we shall see this product is not anymore local in
coordinate space but satisfies the Leibnitz rule with respect to
the symmetric difference operator $\Delta_\pm$. This is easily
checked in the momentum representation. An operation of $\Delta_\pm$
in the coordinate space corresponds in momentum space to multiplication by
$\hat{p}_\pm=\frac{2}{a}\sin\frac{ap_\pm}{2}$, so that the lattice
momentum conservation leads to the Leibnitz rule on the lattice momentum
space:
\beq \hat{p}_\pm~ (F*G)(p)
 = \left(\frac{a}{2\pi}\right)^2\int d^2 \hat{p}_1\, d^2 \hat{p}_2 \left[
\hat{p}_{1\pm} F(p_1) ~
G(p_2)+ F(p_1)\,\hat{p}_{2\pm}G(p_2) \right]
\delta^{(2)}(\hat{p}-\hat{p}_1-\hat{p}_2).   \label{lbrule}
\eeq
It is thus natural to expect that Leibnitz rule will be satisfied on the
corresponding coordinate space where the lattice momentum conservation
is satisfied. On this new coordinate space the product of fields is
modified. We call this new product as a new type of star
$*$-product and it was proposed in our previous paper\cite{D'Adda:2010pg}.
Similar proposal\cite{Nojiri1} was given at the early investigation of lattice
supersymmetry following to the suggestion of the lattice momentum
conservation of Dondi and Nicolai\cite{D-N}.

Explicit form of the coordinate representation of the star product
is given by
\bea
(F*G) (x)
&=& F(x) * G(x)
=a^2\int \frac{d^2\hat{p}}{(2\pi)^2}\, e^{-ipx}~ (F*G)(p)
\nonumber \\
&=&\int_{-\frac{\pi}{2}}^{\frac{3\pi}{2}} d^2\tilde{p}~
\cos\tilde{p}_+\cos\tilde{p}_-
~e^{-ipx} \int_{-\frac{\pi}{2}}^{\frac{3\pi}{2}}
\frac{d^2\tilde{p}_1}{(2\pi)^2} \frac{d^2\tilde{p}_2}{(2\pi)^2}
\cos\tilde{p}_{1+}\cos\tilde{p}_{1-}\, \cos\tilde{p}_{2+}\cos\tilde{p}_{2-}
\nonumber \\
&&\times \int_{-\infty}^\infty  \frac{d^2\tau}{(2\pi)^2}
e^{i{\tau}\cdot(\sin\tilde{p}-\sin\tilde{p}_1-\sin\tilde{p}_2)}
\sum_{y_\pm,z_\pm}e^{i(m\cdot\tilde{p}_1 +l\cdot\tilde{p}_2)}
F(y) G(z)
\nonumber \\
&=& \int_{-\infty}^\infty d^2\tau J_{n_+\pm1}({\tau}_+)J_{n_-\pm1}({\tau}_-)
\nonumber\\
&&\times \sum_{m_\pm,l_\pm}J_{m_+\pm1}({\tau}_+)J_{m_-\pm1}({\tau}_-)
J_{l_+\pm1}(\tau_+)J_{l_-\pm1}(\tau_-) F(y) G(z),
\label{def-star-prod-cor}
\eea
where $\tilde{p}_\pm=\frac{ap_\pm}{2}$, 
and $x_\pm=\frac{n_\pm a}{2}, y_\pm=\frac{m_\pm a}{2}, z_\pm=\frac{l_\pm a}{2}$ should be understood and the integration variable is
not $p_\pm$ but $\hat{p}_\pm$.

The lattice delta function is parameterized by $\tau_\pm$
\beq
\delta^{(2)}\left(\frac{2}{a}\sin\tilde{p}_i\right)
=\left(\frac{a}{4\pi}\right)^2\int^\infty_{-\infty}
d^2\tau e^{i({\tau}_+ \sin\tilde{p}_{i+} + \tau_- \sin\tilde{p}_{i-})}.
\label{deltafunction}
\eeq
$J_n(\tau_\pm)$ is a Bessel function defined as
\begin{align}
J_n(\tau_\pm)=\frac{1}{2\pi}\int_\alpha^{2\pi+\alpha}
e^{i(n\theta-\tau_\pm \sin \theta)} d\theta,
\label{Bessel}
\end{align}
and we use the following notation:
\begin{align}
J_{n\pm1}(\tau)=\frac{1}{2}(J_{n+1}(\tau) + J_{n-1}(\tau)).
\label{def-Jpm}
\end{align}
It is obvious that the star product is commutative:
\beq
F(x) * G(x) = G(x) * F(x).
\label{star-commut}
\eeq
We can now explicitly check how the difference operator acts on the star
product of two lattice fields and find that the difference operator
action on the star product indeed satisfies Leibnitz rule:
\begin{align}
 i\Delta_\pm (F(x)*G(x)) &=
a^2\int \frac{d\hat{p}}{(2\pi)^2}~ i\Delta_{x_\pm}~e^{-ipx}~(F*G)(p)
\nonumber \\
&=\left(\frac{a^2}{4}\right)^2 \int d^2\hat{p}~e^{-ipx}\sum_{y_\pm,z_\pm}\int
\frac{d^2\hat{p}_1}{(2\pi)^2} \frac{d^2\hat{p}_2}{(2\pi)^2}~e^{ip_1y+ip_2z}
\nonumber \\
&~~~~~~~\times \left((i\Delta_{y_\pm}~F(y))G(z)~+~
F(y)~(i\Delta_{z_\pm}~G(z)) \right)
~\delta^{(2)} (\hat{p}-\hat{p}_1-\hat{p}_2)\nonumber \\
&=(i\Delta_\pm F(x))*G(x) + F(x) * (i\Delta_\pm G(x)).
\label{star-Leibniz}
\end{align}

We can now generalize the definition of the star product for n fields as:
\bea
F_1(x+b_1)* F_2(x+b_2) * &\cdots& *F_n(x+b_n) = 
 \int^\infty_{-\infty} d^2\tau
J_{n_+\pm1}(\tau_+)J_{n_-\pm1}(\tau_-)
\label{star-n-prod}\\
&&\sum_{m_1\pm,\cdots,m_n\pm}
\left(\prod_{j=1}^n J_{m_j+\pm1}(\tau_+)J_{m_j-\pm1}(\tau_-)
F_j(y_j+b_j)\right),
\nonumber
\eea
where $\frac{2x_\pm}{a}=n_\pm, \frac{2y_\pm}{a}=m_{j\pm}$.
It should be noted that interaction terms includes the product a
higher powers of lattice fields. As we can see from the above expression
the kernel of an interaction term includes an integration of Bessel
functions which is integrable. Thus a possible nonlocality of interaction
terms is expected to be mild due to the integrability of the kernel.

The kinetic term of $N=2$ lattice Wess-Zumino action in two dimensions
can be written with the star product form as:
\bea
S_K~=~\sum_{x_\pm}\{ -\bar{\phi}(x)*\Delta_+\Delta_-\phi(x) -
\bar{F}(x)*F(x)+\bar{\psi}_2(x) * \Delta_+\psi_2(x)
+\bar{\psi}_1(x) * \Delta_-\psi_1(x) \}
\label{WZ-coord-action}
\eea
The interaction term can be given as
\bea
S_I~=~ \sum_{x_\pm}\{ iF(x)*(\phi(x))^{n-1} +
\psi_2(x)*\psi_1(x)*(\phi(x))^{n-2}  \}.
\eea
These actions are exactly $N=2$ supersymmetry invariant on
the star product since the difference operator satisfies
Leibnitz rule.

\section{ Conclusion and Discussions}
\label{section7}
\setcounter{equation}{0}

In this paper we have proposed a formulation which solves
 two of the major difficulties in the lattice supersymmetry:
(1) the breaking of the Leibnitz rule by the finite difference
operator and (2) the doubling problem of chiral fermions.

In order to solve the problem (1) and yet  keep an
exact lattice supersymmetric formulation, we take a stance
of choosing the sine of the momentum, rather than the momentum itself,
as conserved quantity on the lattice. As  mentioned in the introduction
this agrees with lattice periodicity in momentum space but leads in coordinate space
to a nonlocal interaction terms and to a
loss of finite lattice translational invariance.
The problem (2) is solved by introducing a half lattice
structure and identifying species doublers as super
partners for fermions and bosons.
Since the species doublers are identified as physical
super partners there is essentially no chiral fermion
problem.

The basic structure of the lattice supersymmetry
transformation which we have proposed
here can be seen by a coordinate representation of the
simplest supersymmetric model in one dimension.
In order to accommodate lattice supersymmetry
transformations it is crucial to introduce a lattice with
spacing $\frac{a}{2}$ together with an alternating sign change.
In  momentum space the alternating sign multiplication
corresponds to a momentum shift of $\frac{2\pi}{a}$, namely  to a species doubler state.
Thus the species doubler states are mixed  under the
supersymmetry transformation.

The lattice version of $N=2$ exact supersymmetry algebra
including superderivative in one and two
dimensions are fully reconstructed
especially in the momentum space.
We found algebraic chiral conditions
which are fundamentally connected with species doubling
structure of chiral fermions on the lattice.
In particular the irreducible representations of chiral
and anti-chiral fields turn out to be fully symmetric
or antisymmetric combinations of the original field
and the species doubler field in each momentum direction.
Thus the chiral condition truncates the reducible part
of unnecessary doubler fields. In two dimensional $N=2$
the 16 fields of the complete superfield (including  doublers
of fermion and boson)  are truncated  4  states by the
chiral conditions forming an irreducible representation.
The chiral conditions require to define newly rescaled
fields which have different periodicity from the
original fields in  momentum space. The original
fields are scattered and separated by $\frac{a}{4}$ spacing.
The chiral condition has the effect of defining  new
fields that gather into
the same site of ($\frac{a}{4},\frac{a}{4}$) lattice,
having half lattice spacing into each light cone
direction.

The one dimensional action derived heuristically in the
previous paper \cite{D'Adda:2010pg} is derived as a chiral super charge
exact form of the super fields together with the lattice version of sine
momentum conservation.
On the other hand the two dimensional $N=2$ supersymmetric
Wess-Zumino action can be derived as a chiral super charge
exact form of the chiral and anti-chiral super fields.
It would be interesting to reveal the fundamental difference
of the formulations between one and two dimensions: In one
dimension all the species doublers are used to construct the
action while 1/4 species doublers out of $4\times4$=16 are truncated
into 4 by the chiral condition and 4+4=8 chiral and anti-chiral
fields are needed for the two dimensional Wess-Zumino action.
The invariance under $N=2$ exact lattice supersymmetry
of the actions is assured by the nil-potency of chiral
super charges and by the lattice momentum conservation.

Since the algebra has a lattice periodicity in the
momentum space, it is natural that delta function
should have the same periodicity and thus leading to the
lattice sine momentum conservation. The corresponding
coordinate representation of the formulation defines
a new type of product which we called a star
product\cite{D'Adda:2010pg}.
This product has a nonlocal nature but expected to have
milder non-locality than the non-locality generated by
the SLAC type derivative since the kernel of the
interaction terms are expressed by Bessel functions and
thus they are integrable.
It is important to note that the difference operator
satisfies the Leibnitz rule on star product fields.
Thus exact supersymmetry invariance on the coordinate
space with the star product fields is assured in
parallel to the momentum representation.

The periodic momentum conservation unavoidably enforces
us to introduce a non-locality and non-translational
invariance into the formulation. As we argued in the
introduction the non-locality would be unavoidable in the
formulation if we strictly require the exact lattice
supersymmetry for the formulation.
We consider that the lattice version of sine
momentum conservation is the best possible choice if we
keep strictly an exact lattice supersymmetry.
It is shown that the translational invariance is
recovered in the continuum limit.

In solving lattice chiral fermion problem, exponentially damping
non-locality was accepted and an modification of lattice chiral
transformation was introduced. By breaking one of presumed
conditions of No-Go theorem\cite{No-Go} the satisfactory solution for
the lattice chiral fermion was obtained. We may take a similar
stance for solving the lattice supersymmetry. In other wards
we accept the breakdown of locality and translational invariance
in the lattice supersymmetry formulation and then expect that
the locality and translational invariance recovers harmlessly
in the continuum limit. We have confirmed that this recovery
of the translational invariance is realized even with quantum
corrections from dimensional arguments in the Appendix.
It is, however, important to see if this is the case even in
the nonperturbative regime. We may say that it is similar to expect
the recovery of Lorentz invariance in the continuum limit.
It is also important to investigate carefully if the exactness
of the lattice supersymmetry is kept even at the quantum level.
In our previous paper I\cite{D'Adda:2010pg} we showed that the
lattice supersymmetry
is kept at the one loop quantum level.
It is also necessary to investigate a quantum level exactness
of the lattice supersymmetry given in this paper.
The extension of the formulation given here into
higher dimensions and also gauge theory is obviously important.
The question of feasibility for numerical evaluation may be
important. In this paper we have formulated Minkowsky version of
lattice supersymmetry. It may be neccessary to reformulate in the
Euclidean lattice for numerical simulation.

\vspace{1cm}

{\bf{\Large Acknowledgments}}
 N.K. would like to thank K. Asaka, Y. Kondo, E. Giguere for
useful discussions.
This work was supported in part by Japanese Ministry of Education,
Science, Sports and Culture under the grant number 22540261 and
also by the research funds of Insituto Nazionale di Fisica
Nucleare (INFN). I.K. is supported in part by the Deutsche
Forschungsgemeinschaft
(Sonder-forschungsbereich / Transregio 55) and
the Research Executive Agency (REA) of the European Union
under Grant Agreement number PITN-GA-2009-238353 (ITN STRONGnet).

\begin{center}
{\bf{\Large Appendix}}
\end{center}

\appendix
\section{Periodicity, hermiticity and symmetry properties of lattice fields}

In this appendix we summarize and review the periodicity properties of the lattice fields
in the momentum representations, and how these are related to hermiticity and symmetry properties under
$p_{\pm} \rightarrow \frac{2 \pi}{a} - p_{\pm}$. These properties have been discussed and used through the paper and are collected here for the reader's convenience.

\subsection{Periodicity in $p_{\pm}$}

In the present approach each lattice field $\Phi_A$ is defined on a lattice with spacing $\frac{a}{2}$  defined by
\beq
\vec{x}_\pm = \{ (n+b_A)\frac{a}{2},~ (m+c_A)\frac{a}{2}\}, ~~~~~~n,m\in {\bf Z}, \label{l-con-coora}
\eeq
where $b_A$ and $c_A$ represent the shift of the lattice on which the field $\Phi_A$ is defined with respect to the origin of the coordinate space. In the formulation of chiral lattice supersymmetry given in the present paper it is not possible to set all $b_A$ and $c_A$ to zero, because supersymmetry transformations are defined in terms of symmetric finite differences of spacing $\frac{a}{2}$ as shown in (\ref{susy1+}).
As a result, if we set the shifts $b$ and $c$ equal to zero for the first component $\Phi$ of the superfield, the different fields of the $N=2$ supermultiplet are effectively defined on a lattice with spacing $\frac{a}{4}$, as shown in fig. \ref{fig:field-scattered}.
In the momentum representation a field $\Phi_A$ defined on the lattice (\ref{l-con-coora}) is periodic in the momenta with period $\frac{4\pi}{a}$ up to a phase that depends on the shifts $b_A$ and $c_A$. more precisely we have:
\bea
\Phi_A(p_+ + \frac{4\pi}{a}, p_- )& = & e^{2i\pi b_A} \Phi_A(p_+,p_-) \nonumber \\
\Phi_A(p_+, p_- + \frac{4\pi}{a}) &=& e^{2i\pi c_A} \Phi_A(p_+,p_-)
\label{pe}
\eea
With the choice of shifts corresponding to the lattice of fig. \ref{fig:field-scattered} the phases at the r.h.s. of (\ref{pe}) are just $\pm 1$, and the periodicity conditions of the different field are the ones given in (\ref{period2}), which we reproduce here for convenience and we denote as case A:
\bea
\Phi(p_+,p_-)=&\Phi(p_+ +\frac{4\pi}{a},p_-)=\Phi(p_+,p_- +\frac{4\pi}{a}),\nonumber \\ \fbox{Case A}\qquad \Psi_1(p_+,p_-)=&-\Psi_1(p_+ +\frac{4\pi}{a},p_-)=\Psi_1(p_+,p_- +\frac{4\pi}{a}), \nonumber \\\Psi_2(p_+,p_-)=&\Psi_2(p_+ +\frac{4\pi}{a},p_-)=-\Psi_2(p_+,p_- +\frac{4\pi}{a}), \nonumber \\ F(p_+,p_-)=&-F(p_+ +\frac{4\pi}{a},p_-)=-F(p_+,p_- +\frac{4\pi}{a}),\nonumber \\ \label{period2app}
\eea
The important point here is that while the relative shifts $b_A-b_{A'}$ and $c_A-c_{A'}
$ of the different component fields are determined by the algebraic structure of supersymmetry transformations, the latter allows an arbitrary choice for the origin of the $\frac{a}{4}$ lattice of fig. \ref{fig:field-scattered}. The periodicity conditions (\ref{period2app}) correspond to the lattice origin coinciding with the site of a $\Phi$ field, namely to the first component of the superfield $\Phi(x_+,x_-)$ being defined on the lattice of sites $(\frac{n}{2},\frac{m}{2})$. But other choices are equally consistent with the algebra: in particular we can choose the last component $F$ to be defined on the lattice of sites $(\frac{n}{2},\frac{m}{2})$. This corresponds, with respect to the previous choice, to a shift of the origin of $\frac{a}{4}$ in each light cone direction. In momentum representation this means exchanging periodicity with antiperiodicity in (\ref{period2app}), namely (case B):
\bea
\Phi(p_+,p_-)=&-\Phi(p_+ +\frac{4\pi}{a},p_-)=-\Phi(p_+,p_- +\frac{4\pi}{a}),\nonumber \\ \fbox{Case B}\qquad \Psi_1(p_+,p_-)=&\Psi_1(p_+ +\frac{4\pi}{a},p_-)=-\Psi_1(p_+,p_- +\frac{4\pi}{a}), \nonumber \\\Psi_2(p_+,p_-)=&-\Psi_2(p_+ +\frac{4\pi}{a},p_-)=\Psi_2(p_+,p_- +\frac{4\pi}{a}), \nonumber \\ F(p_+,p_-)=&F(p_+ +\frac{4\pi}{a},p_-)=F(p_+,p_- +\frac{4\pi}{a}),\nonumber \\ \label{period3app}
\eea

In terms of the rescaled fields $\uphi_A$ defined in (\ref{def-chi-fields-2d}), which are the relevant ones for writing the chiral conditions, the periodicity conditions are particularly simple. In coordinate representation all fields $\uphi_A$ are on the same sites of a lattice with spacing $\frac{a}{2}$. In the case A this is given by the sites of coordinates $(\frac{na}{2}+\frac{a}{4},\frac{ma}{2}+\frac{a}{4})$ and is represented in fig. \ref{fig:field-collected}. In the case B the lattice is obtained from case A with a shift of $\frac{a}{4}$ in each light cone direction, it includes the origin and is given by the sites $(\frac{an}{2}, \frac{am}{2})$.
In momentum representation all fields $\uphi_A$ have the same periodicity properties: in case A they are all antiperiodic with period $\frac{4\pi}{a}$ in $p_+$ and $p_-$, in case B on the other hand they are all periodic.

\subsection{Hermiticity and symmetry properties under $p_{\pm} \rightarrow \frac{2\pi}{a} - p_{\pm}$}

We have assumed all through the paper that the fields $\Phi_A(x)$ in coordinate representation are real.
In momentum representation this means:
\beq
\Phi^\dagger_A(p) = \Phi_A(-p).
\label{herm2}
\eeq
On the other hand we discussed in Section 4 that the chiral and anti-chiral parts of the  component fields $\uphi_A$ are given respectively by the symmetric and antisymmetric parts with respect to the symmetry transformations $p_\pm \rightarrow \frac{2\pi}{a} - p_\pm$ as shown in (\ref{chiral2}) and (\ref{antichiral}).
We want to show now how the hermiticity properties of chiral and antichiral fields depend on the choice of the periodicity properties ( case A and B of previous subsection). Let us concentrate on the dependence of the fields  on say the $p_+$ variable ( $p_-$ remains a spectator and will be ignored although of course the same considerations can be repeated for $p_-$). Consider the symmetric and antisymmetric part of $\uphi_A(p_+)$ with respect to $p_+ \rightarrow \frac{2\pi}{a}-p_+$:
  \beq
  \uphi^{(\pm)}_A(p_+) = \frac{1}{2} \left( \uphi_A(p_+) \pm \uphi_A(\frac{2\pi}{a} - p_+) \right)
  \label{sya}
  \eeq
  If one takes the hermitian conjugate of say $\uphi^{(+)}_A(p_+)$ one gets, using (\ref{herm2}):
  \beq
  \uphi^{(+)\dagger}_A(p_+) = \frac{1}{2} \left( \uphi_A(-p_+) \pm \uphi_A(-\frac{2\pi}{a} + p_+) \right)
  \label{a4}
  \eeq
  If one chooses the periodicity conditions labeled as A, namely all rescaled fields $\uphi_A$ antiperiodic, we obtain from (\ref{a4}):
  \beq
  \uphi^{(+)\dagger}_A(p_+) = \uphi^{(-)}_A(-p_+),
  \label{a5}
  \eeq
  namely the chiral and antichiral fields are hermitian conjugate of each other. On the other hand, with the choice B ( all $\uphi_A$ periodic ) we have:
  \beq
  \uphi^{(+)\dagger}_A(p_+) = \uphi^{(+)}_A(-p_+),
  \label{a5b}
  \eeq
  namely chiral and antichiral fields fields are real.

  In the previous considerations chiral and antichiral fields are thought of as part of the full superfield expansion, and in fact arise from decomposing the superfield into irreducible representations of the supersymmetry algebra. In this context chiral fields are symmetric with respect to $p_\pm \rightarrow \frac{2\pi}{a} - p_\pm$ and antichiral fields antisymmetric.
  However they can be considered by themselves as irreducible representations of the supersymmetry algebra, and in that case they are only constrained by the requirement of satisfying the supersymmetry transformations given in Tables 2 and 3. These transformations are consistent with the fields
  $\uphi_A$ being either all periodic or all antiperiodic in $p_\pm$ with period $\frac{4\pi}{a}$. They are also consistent with all fields being symmetric or all antisymmetric with respect to $p_\pm \rightarrow \frac{2\pi}{a} - p_\pm$. This means that chiral and antichiral superfields do not need to have opposite symmetry properties with respect to such transformation if they are taken as the  fundamental building blocks of the theory, rather than resulting from the decomposition of the full superfield representation.
  In Section 5 this freedom proved to be essential in constructing an action with the right symmetry properties.  As discussed in section 5 it also allows the chiral and antichiral fields to be  hermitian conjugate ( see eq. (\ref{herm})) and at the same time periodic in $p_\pm$.

\section{Recovery of the translational invariance}
\label{sec:trans-inv}

In this appendix, we show that the translational invariance of $n$-point
correlation function is recovered in the continuum limit.

Let us denote the component fields as $\phi_A=(\varphi, \psi_1,\psi_2, f,
\bar{\varphi}, \bar{\psi}_1, \bar{\psi}_2, \bar{f})$.
Because of the sine momentum conservation law correlation functions are
invariant under the following transformation:
\begin{equation}
 \phi_A(p_+,p_-)
 \to \exp\left(i\sum_{i=+,-} l_i \frac{2}{a}\sin\frac{ap_i}{2} \right)
     \phi_A(p_+,p_-)
 \label{eq:sine-translation}
\end{equation}
with finite length $l_+, l_-$, whereas an invariance under finite
translation should be the invariance under the transformation:
\begin{equation}
 \phi_A(p_+, p_-)
 \to \exp(i \sum_{i=+,-} l_i p_i) \phi_A(p_+,p_-).
\end{equation}
Under transformation (\ref{eq:sine-translation}) $n$-point correlation
function transforms as
\begin{align}
 \lefteqn{
  \langle \phi_{A_1}(p_{1+}, p_{1-})
   \phi_{A_2}(p_{2+}, p_{2-})\cdots \phi_{A_n}(p_{n+}, p_{n-})
 \rangle
 }\qquad \nonumber \\
&\to \exp(i \sum_{I=1}^n \sum_{i=+,-} l_i \frac{2}{a}\sin\frac{ap_{Ii}}{2})
  \langle \phi_{A_1}(p_{1+}, p_{1-})
   \phi_{A_2}(p_{2+}, p_{2-})\cdots \phi_{A_n}(p_{n+}, p_{n-})
 \rangle \nonumber \\
&\simeq
  \exp(i \sum_{I=1}^n \sum_{i=+,-} l_i p_{Ii})
  \left[1
 -i\frac{la^2}{24} \left(\sum_{I=1}^n \sum_{i=+,-}p_{Ii}\right)^3 \right]
  \\
&\langle \phi_{A_1}(p_{1+}, p_{1-})
   \phi_{A_2}(p_{2+}, p_{2-})\cdots \phi_{A_n}(p_{n+}, p_{n-})
 \rangle, \nonumber
\end{align}
which shows that if the correlation function is less divergent than
$1/a^2$ the translational invariance is recovered in the continuum limit.

Now let us check the divergences of the correlation
functions.  Because of the mass term there is no infrared divergence
so we concentrate on the ultraviolet divergences.

After rescaling the fields and couplings having dimensions,
the interaction terms do not contain any divergences.  The quadratic terms,
i.e., the kinetic term and mass term are given
by\footnote{We also have rescaled the overalll factor of the action.}
\begin{align}
 S_2
 &=
  \int_{\pi/a}^{\pi/a}dp_+ dp_- \Bigl[
   -\bar{\varphi}(-p)\frac{4}{a^2}\sin\frac{ap_+}{2}\sin\frac{ap_-}{2}
\varphi(p)
   -\bar{f}(-p)f(p)
 \nonumber\\
   &-im\bar{\varphi}(-p)\bar{f}(p)
   +im f(-p) \varphi(p) 
  + \bar{\psi}_1(-p)\frac{2}{a}\sin\frac{ap_-}{2}\psi_1(p)
\nonumber\\
  &+ \bar{\psi}_2(-p)\frac{2}{a}\sin\frac{ap_+}{2}\psi_2(p)
  + m \psi_1(-p)\psi_2(p)
  - m \bar{\psi}_1(-p)\bar{\psi}_2(p)
\Bigr],
\end{align}
leading to the following propagators and ultraviolet behaviors:
\begin{align}
 \langle \varphi(p)\bar{\varphi}(-p)\rangle
   &= \frac{-1}{D(p)}  \sim a^2\\
 \langle f(p) \bar{f}(-p) \rangle
  &= \frac{-1}{D(p)}\frac{4}{a^2}\sin\frac{ap_+}{2}\sin\frac{ap_-}{2}
   \sim a^0 \\
 \langle \varphi(p)f(-p)\rangle
  &=  -\langle \bar{\varphi(p)}\bar{f(-p)}\rangle
  = \frac{-im}{D(p)}  \sim a^2 \\
 \langle \psi_1(p)\bar{\psi}_1(-p) \rangle
  &= \frac{1}{D(p)}\frac{2}{a}\sin\frac{ap_+}{2}  \sim a \\
 \langle \psi_2(p)\bar{\psi}_2(-p) \rangle
  &= \frac{1}{D(p)}\frac{2}{a}\sin\frac{ap_-}{2}  \sim a \\
 \langle \psi_1(p)\psi_2(-p) \rangle
  &= - \langle \bar{\psi}_1(p)\bar{\psi}_2(-p) \rangle
  =\frac{m}{D(p)}  \sim a^2,
\end{align}
where
\begin{equation}
 D(p)=\frac{4}{a^2}\sin\frac{ap_+}{2}\sin\frac{ap_-}{2} -m^2.
\end{equation}
Consider now an amputated diagram with $I_B^{f\bar{f}}$ internal
$f\bar{f}$ lines, $I_F^{\rm diag.}$ internal $\psi_1 \bar{\psi}_1$
and $\psi_2 \bar{\psi}_2$ lines, and $V$ vertexes.
Taking into account $1/a^2$ from each of loop momentum integrations,
and $a^2$ from momentum conservation at each vertexes which reduces
number of integrations (except for the total momentum conservation
of external lines), we obtain
\begin{equation}
 (\text{the total UV contribution})
  \sim \left(\frac{1}{a^2}\right)^{I^{f\bar{f}}}
       \left(\frac{1}{a}\right)^{I_F^{\rm diag.}}
       (a^2)^{V-1}.
\end{equation}
The number of vertexes $V$ is decomposed into two parts, $V=V_B+V_F$,
one is a number of bosonic vertexes $V_B$ and the other is that of
fermionic vertexes $V_F$.
They satisfy the following relations:
\begin{align}
 V_B&= 2I_B^{f\bar{f}} + E_B^f,\\
 V_F&= I_F^{\rm diag.} + I_F^{\text{off-diag.}} + \frac{1}{2}E_F,
\end{align}
where $I_F^{\text{off-diag.}}$ is number of internal fermion lines of
$\psi_1\psi_2$ and $\bar{\psi}_1\bar{\psi}_2$, and $E_B^f$ and $E_F$
are  number of external $f$ and fermions.
Combining above relations, we obtain
\begin{equation}
 (\text{the total UV contribution})
  \sim a^{I_F^{\rm diag.}+2I_F^{\text{off-diag.}} + 2I_B^{f\bar{f}}+E_B^f + \frac{1}{2}E_F -2}.
\end{equation}
The most divergent case would be
\begin{equation}
  I_F^{\rm diag.}=I_F^{\text{off-diag.}} =I_B^{f\bar{f}} = E_B^f = E_F =0
\end{equation}
which gives $1/a^2$. We cannot make, however, a loop in this case.
Therefore the UV divergence is less than $1/a^2$ and thus the
translational invariance is recovered in the continuum limit.



\begin{thebibliography}{99}






















\bibitem{Montvay}
I.~Montvay,
Int.\ J.\ Mod.\ Phys.\ {\bf A17} (2002) 2377 [arXiv:hep-lat/0112007].

\bibitem{Feo:2002yi}
 A.~Feo,
 Nucl.\ Phys.\ Proc.\ Suppl.\  {\bf 119} (2003) 198
  [arXiv:hep-lat/0210015].

\bibitem{Kaplan:2003uh}
D.~B.~Kaplan,
Nucl.\ Phys.\ Proc.\ Suppl.\  {\bf 129} (2004) 109
[arXiv:hep-lat/0309099].


\bibitem{Catterall:2004np}
  S.~Catterall,
  JHEP {\bf 0411} (2004) 006
  [arXiv:hep-lat/0410052].


\bibitem{Giedt:1}
J.~Giedt,
PoS{\bf LAT2006} (2006) 008
[{ arXiv:hep-lat/0701006}].

\bibitem{CKU}
S.~Catterall, D.~Kaplan and M.~Unsal,
Phys.\ Rep.\ 484 (2009) 71 [arXiv:hep-lat/0903.4881].

\bibitem{D-N}
  P.~H.~Dondi and H.~Nicolai,
  Nuovo Cim.\  A {\bf 41} (1977) 1.

\bibitem{chiral-fermion}
P.~Hasenfratz,
Nucl.\ Phys.{\bf(Proc. Suppl.) 63A-C} (1998) 53 [arXiv:hep-lat/9709110]. \\
H.~Neuberger,
Phys.\ Lett. {\bf B427} (1998) 353 [arXiv:hep-lat/9801031]. \\
M.~L\"uscher,
Phys.\ Lett. {\bf B428} (1998) 342 [arXiv:hep-lat/9802011].

\bibitem{Fujikawa:Leibniz}
K.~Fujikawa,
Nucl.\ Phys.\  B {\bf 636} (2002) 80
[arXiv:hep-th/0205095].

\bibitem{C-G}
S.~Catterall and E.~Gregory,
Phys.\ Lett.\ B {\bf 487} (2000) 349 [arXiv:hep-lat/0006013].

\bibitem{Giedt}
J.~Giedt, R.~Koniuk, E.~Poppitz and T.~Yavin,
JHEP 12 (2004) 033 [arXiv:hep-lat/0006013].

\bibitem{Bergner:2009vg}
  G.~Bergner,
  JHEP {\bf 1001} (2010) 024
  [arXiv:0909.4791 [hep-lat]].

\bibitem{KBUWW}
T.~Kaestner, G.~Bergner, S.~Uhlmann, A.~Wipf, and C.~Wozar
Phys. Rev. {\bf D78} (2008) 095001 [arXiv:0807.1905 [hep-lat] ].

\bibitem{Kato:2008sp}
  M.~Kato, M.~Sakamoto and H.~So,
  JHEP {\bf 0805} (2008) 057
  [arXiv:0803.3121 [hep-lat]].

\bibitem{KatoSakamotoSo}
 M.~ Kato, M.~ Sakamoto and H.~ So,
  PoS LAT2005:274 (2006) [hep-lat/0509149]; PoS LATTICE2008:233 (2008)
	[arXiv:0810.2360 [hep-lat] ].

\bibitem{Nojiri1}
 S.~Nojiri,
 Prog. Theor. Phys. {\bf 74} (1985) 819,
  ibid {\bf 74} (1985) 1124.

\bibitem{BartelsKramer}
J.~Bartels ad G.~Kramer,
Z.Phys. {\bf C20} (1983) 159.

\bibitem{DKKN}
  A.~D'Adda, I.~Kanamori, N.~Kawamoto and K.~Nagata,
  Nucl.\ Phys.\ {\bf B707} (2005) 100
  [arXiv:hep-lat/0406029],\\
  A.~D'Adda, I.~Kanamori, N.~Kawamoto and K.~Nagata,
  Phys.\ Lett.\  B {\bf 633} (2006) 645
  [arXiv:hep-lat/0507029],\\
  A.~D'Adda, I.~Kanamori, N.~Kawamoto and K.~Nagata,
  Nucl.\ Phys.\  B {\bf 798} (2008) 168
  [arXiv:0707.3533 [hep-lat]].

\bibitem{SLAC-der}
S.~D.~Drell, M.~Weinstein, and S.~Yankielowicz,
Phys. Rev. {\bf D14} (1976) 1627.

\bibitem{D'Adda:2010pg}
  A.~D'Adda, A.~Feo, I.~Kanamori, N.~Kawamoto and J.~Saito,
  JHEP {\bf 1009} (2010) 059
  [arXiv:1006.2046 [hep-lat]].


\bibitem{K-S2}
L.~H.~Karsten and J.~Smit,
Phys. Lett. {\bf B85} (1979) 100.

\bibitem{BBP}
G.~Bergner, F.~Bruckmann, and J.~M.~ Pawlowski,
Phys. Rev. {\bf D79} (2009) 115007 [arXiv:0807.1110 [hep-lat]].

\bibitem{Dutch}
  F.~Bruckmann and M.~de Kok,
  Phys.\ Rev.\  D {\bf 73} (2006) 074511
  [arXiv:hep-lat/0603003].

\bibitem{B-C-K}
  F.~Bruckmann, S.~Catterall and M.~de Kok,
  Phys.\ Rev.\  D {\bf 75} (2007) 045016
  [arXiv:hep-lat/0611001].

\bibitem{D'Adda:2009jb}
  A.~D'Adda, N.~Kawamoto and J.~Saito,
  PoS {\bf LAT2009} (2009) 047
  [arXiv:0910.3149 [hep-lat]].



\bibitem{old-Dirac-Kahler}
S.~Elitzur, E.~Rabinovici and A.~Schwimmer,
Phys. Lett. {\bf B 199} (1982) 165.

\bibitem{Kaplan:2003}
D.~B.~Kaplan, E.~Katz and M.~\"{U}nsal,
JHEP {\bf 0305} (2003) 037
[{ arXiv:hep-lat/0206019}].
\bibitem{Cohen:2003xe}
  A.~G.~Cohen, D.~B.~Kaplan, E.~Katz and M.~Unsal,
  JHEP {\bf 0308} (2003) 024
  [arXiv:hep-lat/0302017];
  JHEP {\bf 0312} (2003) 031
  [arXiv:hep-lat/0307012] .

\bibitem{catterall}
S. Catterall, JHEP {\bf 11} (2004) 006
[arXiv:hep-lat/0410052].

\bibitem{sugino}
F. Sugino, JHEP {\bf 01} (2004) 015
[arXiv:hep-lat/0311021].

\bibitem{D-M}
  P.~H.~Damgaard and S.~Matsuura,
  JHEP
 {\bf 0709} (2007) 097
  [arXiv:0708.4129 [hep-lat]];
  Phys.\ Lett.\  B {\bf 661} (2008) 52
  [arXiv:0801.2936 [hep-th]].

\bibitem{Takimi:2007nn}
  T.~Takimi,
  JHEP {\bf 0707} (2007) 010
  [arXiv:0705.3831 [hep-lat]].

\bibitem{Arianos:2008ai}
S.~Arianos, A.~D'Adda, A.~Feo, N.~Kawamoto and J.~Saito,
Int.\ J.\ Mod.\ Phys. {\bf A24} (2009) 4737
[arXiv:hep/lat/0806.0686].

\bibitem{D'Adda:2009es}
  A.~D'Adda, A.~Feo, I.~Kanamori, N.~Kawamoto and J.~Saito,
  PoS {\bf LAT2009} (2010) 051
  [arXiv:0910.2924 [hep-lat]].

\bibitem{KKU}
 N.~Kawamoto and T.~Tsukioka,
  Phys.\ Rev.\  D {\bf 61} (2000) 105009
  [arXiv:hep-th/9905222];\\
  J.~Kato, N.~Kawamoto and Y.~Uchida,
  Int.\ J.\ Mod.\ Phys.\  A {\bf 19} (2004) 2149
  [arXiv:hep-th/0310242];\\
  J.~Kato, N.~Kawamoto and A.~Miyake,
  Nucl.\ Phys.\  B {\bf 721} (2005) 229
  [arXiv:hep-th/0502119].

\bibitem{No-Go}
H.B.Nielsen and M.Ninomiya,
       Nucl.Phys.{\bf B185} (1981) 20,\\
L.H.Karsten and J.Smit,
       Nucl.Phys.{\bf B183} (1981) 103.

\bibitem{K-S}
     N.Kawamoto and J.Smit,
       Nucl.Phys.{\bf B192} (1981) 100.













\end{thebibliography}
\end{document}